%% file: manuscript.tex
\documentclass[
 preprint,
 nofootinbib,
 amsmath,amssymb,
 aps,
 prd,
 a4paper
]{revtex4-2}
\pdfoutput=1

\usepackage{graphicx}
\usepackage{dcolumn}
\usepackage{bm}
\usepackage{slashed}
\usepackage[dvipsnames]{xcolor}
\usepackage{physics}
\usepackage{comment}
\usepackage[normalem]{ulem}
\usepackage{url}
\usepackage{bigints}
\usepackage{hyperref}

\usepackage{tikz}
\usetikzlibrary{arrows.meta,calc,patterns}
\usepackage{standalone}
\newcommand{\figspath}{figs}

\DeclareMathOperator{\U}{U}
\DeclareMathOperator{\SU}{SU}

\begin{document}

\preprint{RIKEN-iTHEMS-Report-26}

\title{Metastable vortex string decay in a brane picture}

\author{Akifumi Chitose}
\email{achitose@icrr.u-tokyo.ac.jp}
\affiliation{ICRR, The University of Tokyo, Kashiwa, Chiba 277-8582, Japan}
\author{Hajime Fukuda}
\email{hfukuda@hep-th.phys.s.u-tokyo.ac.jp}
\affiliation{Department of Physics, The University of Tokyo, Tokyo 113-0033, Japan}
\author{Kantaro Ohmori}
\email{kantaro.omori@riken.jp}
\affiliation{RIKEN iTHEMS, RIKEN, Wako, Saitama 351-0198, Japan}

\date{\today}

\begin{abstract}
We describe the decay of a metastable string as a bounce in a brane configuration.
We consider a particular $\mathcal N=1$ supersymmetric $\U(2)$ gauge theory with cascade Higgsing $\U(2)\to \U(1)^2 \to \U(1)$, where the vortex string can break by creating a pair of 't~Hooft-Polyakov monopole and anti-monopole.
The theory is engineered in a Type IIA brane setup, where the metastable string is realized as a D2-brane stretched between NS5- and D4-branes. The string breaks when the endpoints of the D2-brane, moving along the NS5- and D4-branes, meet so that the D2-brane shrinks to nothing.
We check that the tension of the D2-brane reproduces that of the vortex string obtained numerically in the field theory, with the ratio approaching one in the limit where the vortex string becomes BPS.
Within the thin-wall approximation, we derive the bounce action from the Dirac-Born-Infeld action of the D2-brane and find that it agrees with that of the field theory, with the monopole mass given by the area swept by the D2-brane.
As a consistency check, we numerically construct a bounce solution of the D2-brane equation of motion under simplifying assumptions, whose action agrees with the thin-wall result in the regime where our description is reliable.
\end{abstract}

\maketitle
\newpage

\section{Introduction}
\label{sec:intro}
Solitons play important roles in high-energy physics, both from theoretical and phenomenological points of view. For the theoretical point of view, one of the most important solitons is branes in string theory\,\cite{Polchinski:1996na}. Branes are non-perturbative objects in string theory, and they are crucial for understanding not only the non-perturbative dynamics of string theory but also the non-perturbative dynamics of gauge theories realized on the branes\,\cite{Hanany:1996ie,Giveon:1998sr}. For the phenomenological point of view, topological defects in field theories,
such as cosmic strings can be produced in the early universe and can have various cosmological implications\,\cite{Vilenkin:1984ib}.
In particular, cosmic strings can produce gravitational waves, and the recent evidence of gravitational wave background by multiple pulsar timing array (PTA) collaborations\,\cite{Xu:2023wog, NANOGrav:2023gor, NANOGrav:2023hde, NANOGrav:2023hvm, EPTA:2023fyk, Miles:2024seg, Reardon:2023gzh} has triggered a lot of interest in metastable cosmic strings as a possible source of the observed gravitational wave background\,\cite{Buchmuller:2023aus, Afzal:2023kqs, Ahmed:2023pjl, Afzal:2023cyp, Ahmed:2023rky, Lazarides:2023rqf, Lazarides:2023ksx, Antusch:2024ypp, Chitose:2024pmz, Pallis:2024joc, Maji:2024cwv, Datta:2024bqp, Antusch:2024nqg, Antusch:2024qpb, Pallis:2024mip, Ingoldby:2025wcl, Chitose:2025qyt, Maji:2025thf, Antusch:2025xrs, Hu:2025sxv, Ahmad:2025dds, Blasi:2026iyq, Asl:2026zpj, Tranchedone:2026lav,deGiorgi:2026fyx,Maji:2026bvb,Datta:2026hka,Rybak:2026mbk}.

Interestingly, some topological defects in field theories can be described by branes in string theory\,\cite{Hanany:1997hr,Hanany:2003hp,Hanany:2004ea,Tong:2003pz,Tong:2005un}. For instance, in $\mathcal N = 2$ supersymmetric gauge theories with $\U(N)$ gauge group with $N$ fundamental hypermultiplets, the gauge symmetry is completely broken if we introduce the Fayet-Iliopoulos (FI) term, and vortex strings can exist in the theory as $\pi_1(\U(N)) = \mathbb{Z}$ topological defects. These vortex strings preserve half of the supersymmetry and are Bogomol'nyi-Prasad-Sommerfield (BPS) states. In the brane picture, the $\mathcal N = 2$ supersymmetric gauge theory with $\U(N)$ gauge group with $N$ fundamental hypermultiplets and the FI term is realized by the Hanany-Witten setup\,\cite{Hanany:1996ie,Giveon:1998sr} in Type IIA string theory, which consists of NS5-branes, D4-branes and D6-branes. In this brane configuration, the vortex strings correspond to D2-branes stretched between the NS5-brane and the D4-branes. The tension of the vortex strings can be calculated from the tension of the D2-branes, and it matches with that calculated in the field theory picture. The brane picture also successfully describes the structure of the moduli space of the vortex strings and the dynamics on the string worldsheet. 

Then, a natural question arises: can we describe non-BPS topological defects in field theories using the brane picture as well? There has been several attempts for that. Of the most well-known examples is Sen's conjecture on the tachyon condensation of non-BPS D-branes\,\cite{Sen:1999mg}, where, for example, the unstable pair of a D$(p+2)$-brane and an anti-D$(p+2)$-brane can annihilate into a D$p$-brane, or a kink on an unstable D$(p+1)$-brane can be identified with a D$p$-brane. 
Furthermore, several works have focused on the non-BPS properties of topological defects in field theories and their brane realizations\,\cite{Hanany:1997hr}.

Of particular interest are metastable cosmic strings, which can be produced in relatively simple field theories; suppose, for example, we have a (non-supersymmetric) $\SU(2)$ gauge theory with an adjoint scalar field and a fundamental scalar field in the UV theory and the gauge symmetry is broken hierarchically as $\SU(2) \to \U(1) \to \text{nothing}$. Then, from the second symmetry breaking, we have cosmic strings as $\pi_1(\U(1)) = \mathbb{Z}$, but the cosmic strings are not stable since the vacuum manifold is simply connected: $\pi_1(\SU(2)) = 0$. Therefore, the cosmic strings are metastable with endpoints\,\cite{Vilenkin:1982hm,Preskill:1992ck}, which correspond to monopoles for $\pi_2(\SU(2)/\U(1)) = \mathbb{Z}$.

For hierarchical symmetry breaking, the decay rate of the metastable strings is analytically calculable using the thin-wall approximation\,\cite{Vilenkin:1982hm,Preskill:1992ck}. However, it becomes unreliable in the parameter region of interest for the PTA observations, and one must resort to numerical computations in the field theory\,\cite{Shifman:2002yi,Chitose:2023dam,Domcke:2026xza}.

These metastable strings and their monopole endpoints have also been realized in the brane picture. In Ref.\,\cite{Auzzi:2004yg}, monopoles confined by vortices are realized in MQCD as D2-branes bounded by D4-branes and NS5-branes, and the vortex is identified with a strip of the D2-brane continuing into the spacetime directions, whose width is proportional to the vortex tension. In Ref.\,\cite{Eto:2012kb}, an electroweak-like string and the Nambu monopole at its end are realized by a single D2-brane, and the monopole masses are obtained from the minimal surface area of the D2-brane in the internal space. There, it is also pointed out that the string can break by bending the D2-brane into the internal space, which creates a monopole-anti-monopole pair. However, the tunneling process of this string breaking itself is not described there in the brane picture; for the decay rate, Ref.\,\cite{Eto:2012kb} refers to the thin-wall formula in the field theory\,\cite{Preskill:1992ck}, which is not derived from the dynamics of the D2-brane.

On the other hand, describing tunneling processes by a Euclidean string worldsheet or brane worldvolume that ends on other branes has a long history. The nucleation of brane bubbles bounded by brane junctions was studied in Ref.\,\cite{Gorsky:1999gk}, and the Schwinger-type creation of charge and monopole pairs was interpreted in terms of brane configurations in Ref.\,\cite{Gorsky:2001up}. The pair creation of W-bosons and of open strings has been computed with Euclidean worldsheet instantons, both in flat space\,\cite{Schubert:2010tx,Bolognesi:2016qme} and holographically\,\cite{Semenoff:2011ng,Bolognesi:2016qme}. The breaking of a confining string by quark pair creation, which describes meson decays, has been studied with a worldsheet instanton that splits the string on a flavour brane\,\cite{Peeters:2005fq,Peeters:2018xei}. Decay rates have also been computed in brane configurations for the decay of metastable vacua seeded by solitons\,\cite{Kasai:2015exa,Kasai:2015dia}, where in Ref.\,\cite{Kasai:2015dia} the seed is a D2-brane string identified in Ref.\,\cite{Eto:2006yv}. However, the breaking of a string by monopole-anti-monopole pair creation, where the string itself is a D2-brane ending on other branes, has not been described in this way.

In this paper, we describe this tunneling process in the brane picture and derive its bounce action from the dynamics of the D2-brane.
We focus on the Hanany-Witten setup\,\cite{Hanany:1996ie,Giveon:1998sr} in Type IIA string theory.
There, the $\mathcal N = 1$ supersymmetric gauge theory with $\U(2)$ gauge group is realized with an adjoint chiral multiplet and two fundamental chiral multiplets. The theory undergoes two-step symmetry breaking, $\U(2) \to \U(1)\times \U(1) \to \U(1)$, so that metastable vortex strings can exist associated to the second symmetry breaking. In the brane picture, the metastable vortex strings correspond to non-BPS D2-branes stretched between the NS5-brane and the D4-branes. We first identify the parameters of the field theory with the geometrical parameters of the brane configuration, including the numerical coefficients, and check the identification by comparing the tension of the D2-brane with that of the vortex string obtained by explicitly solving the field equations.

We then describe the string breaking as the tunneling of the D2-brane, and formulate its bounce using the Dirac-Born-Infeld (DBI) action of the Euclidean D2-brane, together with the boundary condition given by the NS5-branes and the D4-branes in the Hanany-Witten setup. The ``stable'' vacuum in this process corresponds to the absence of the metastable string, i.e., the D2-brane itself, where the DBI action is no more applicable. We point out that special care is needed for the bounce there, unlike the tunneling process in field theory\,\cite{Coleman:1977py,Callan:1977pt}: the D2-brane suddenly starts to exist at a finite radius in the Euclidean spacetime. Within the thin-wall approximation, the DBI action of the bounce takes the same form as the field-theory bounce action, $2\pi m R - \pi T R^2$: the D2-brane tension gives the string tension $T$ and the area swept by the D2-brane in the internal space gives the monopole mass $m$, with the string length and the string coupling entering only through the gauge coupling.
Finally, we construct the bounce solution explicitly by numerically solving the Euclidean equation of motion of the D2-brane. We find that such a solution exists and that its action agrees with the thin-wall result, which provides a consistency check of our picture of the tunneling. Since the D2-brane is described by the low-energy DBI action as a thin object, it does not capture the massive modes that form the cores of the monopole and the vortex string; our description is therefore reliable where the thin-wall approximation is valid, i.e., where the bubble is much larger than these cores.

The rest of this paper is organized as follows. In Sec.\,\ref{sec:brane_construction}, we introduce the brane construction of the $\mathcal N = 1$ supersymmetric gauge theory with $\U(2)$ gauge group with an adjoint chiral multiplet and two fundamental chiral multiplets, identify the parameters in the field theory picture with the geometrical parameters in the brane setup, and compare the tension of the D2-brane with that of the vortex string obtained in Appendix\,\ref{sec:vortex_solution}; we also discuss the conditions under which the identification is valid. In Sec.\,\ref{sec:string_breaking}, we describe the string breaking as the tunneling of the D2-brane, formulate its bounce with the DBI action of the Euclidean D2-brane, and show that the bounce action reproduces that of the field theory within the thin-wall approximation. In Sec.\,\ref{sec:numerical_calculation}, we present our numerical algorithm to construct the bounce solution of the Euclidean D2-brane, compare the resulting bounce action with the thin-wall result, and discuss the validity of the thin-wall approximation in the brane and field theory pictures. Sec.\,\ref{sec:conclusion_and_discussion} is devoted to the conclusion and discussion.

\section{Brane construction and parameters}
\label{sec:brane_construction}
In this section, we introduce the brane construction of the $\mathcal N = 1$ supersymmetric gauge theory with $\U(2)$ gauge group with an adjoint chiral multiplet and two fundamental chiral multiplets. As we discuss later, this theory admits a cascade pattern of symmetry breaking, $\U(2) \to \U(1)\times \U(1) \to \U(1)$ in the field theory picture, so that metastable (non-BPS) vortex strings can exist in the intermediate phase with $\U(1)\times \U(1)$ gauge symmetry whereas the metastable vortex strings correspond to non-BPS D2-branes in the brane picture. To perform quantitative analysis of the metastable vortex strings in the brane picture, we need to identify the parameters in the field theory picture with the geometrical parameters in the brane setup with correct numerical coefficients. In this section, we perform this identification and discuss the conditions for the validity of our identification. 

\begin{figure}[t]
\begin{center}
\includegraphics[scale=0.7]{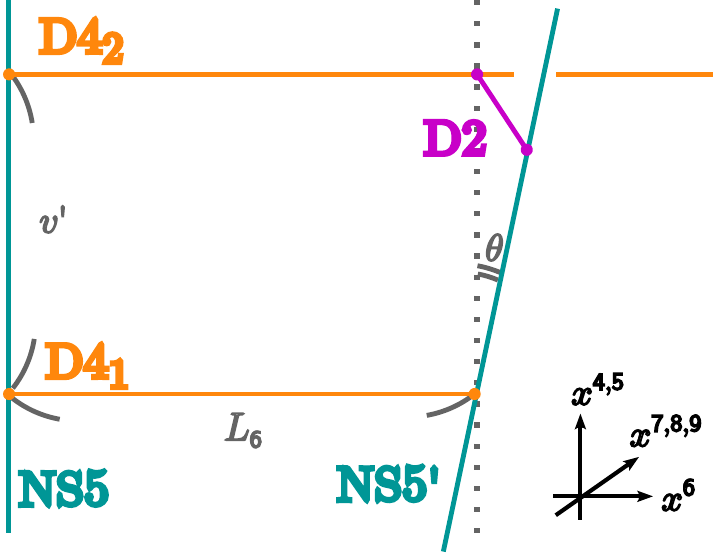}
\end{center}
\caption{
    The brane configuration in Type IIA string theory realizing the $\mathcal N = 1$ supersymmetric gauge theory with $\U(2)$ gauge group with an adjoint chiral multiplet and two fundamental chiral multiplets. The two NS5-branes are separated in the $x^6$ direction by a distance $L_6$. One of the NS5-branes, which we call NS5$'$-brane, is rotated in the $(v, w)$ plane by an angle $\theta$ to break the supersymmetry down to $\mathcal N = 1$. 
    The D4${}_1$-brane is hanging between the two NS5-branes, whereas the D4${}_2$-brane is semi-infinite, ending on the NS5-brane on one side. The distance between the D4${}_1$-brane and the D4${}_2$-brane in the $(x^4, x^5)$ plane is $|v'|$. We also show a (non-BPS) D2-brane, which corresponds to a (non-BPS) vortex string in the field theory picture.
}
\label{fig:hanany-witten}
\end{figure}

The brane configuration we consider here is the Hanany-Witten setup\,\cite{Hanany:1996ie,Giveon:1998sr} in Type IIA string theory, which consists of NS5-brane, NS5$'$-brane, D4${}_1$-branes and D4${}_2$-brane, as shown in Fig.\,\ref{fig:hanany-witten}\,\footnote{Our brane configuration is different from Ref.\,\cite{Eto:2012kb}, where the electroweak symmetry breaking $\SU(2) \times \U(1) \to \U(1)$ is realized with tilted D4-branes and the metastable string is the electroweak $Z$-string ending on the Nambu monopole. Here, instead, we consider the cascade breaking $\U(2) \to \U(1)\times \U(1) \to \U(1)$ with a rotated NS5-brane, so that the endpoint of the metastable string is the 't~Hooft-Polyakov monopole.}.
The NS5-branes are extended along the $x^1, x^2, x^3, x^4, x^5$ directions. One of the NS5-branes, which we call NS5$'$-brane, is rotated to break the supersymmetry down to $\mathcal N = 1$.
The D4${}_1$-branes are suspended between the two NS5-branes along the $x^6$ direction, and the D4${}_2$-brane is semi-infinite, ending on one of the NS5-branes and extending to $x^6 \to +\infty$. The D4-branes are extended along the $x^1, x^2, x^3, x^6$ directions. The $3 + 1$ dimensional gauge theory lives on the $x^1, x^2, x^3$ directions of the branes.
In the language of the field theory picture, this brane configuration realizes the $\mathcal N = 1$ supersymmetric gauge theory with $\U(2)$ gauge group with an adjoint chiral multiplet, $\Phi$, and two fundamental chiral multiplets, $Q$ and $\tilde{Q}$. When the NS5$'$-brane is not rotated, the brane configuration preserves the $\mathcal N = 2$ supersymmetry, so that the adjoint chiral multiplet $\Phi$ is massless and forms the $\mathcal N = 2$ vector multiplet together with the gauge field and the chiral multiplets $Q$ and $\tilde{Q}$ form the $\mathcal N = 2$ hypermultiplet. The superpotential of the theory is given by
\begin{align}
    \label{eq:superpotential_full}
    W = y \tilde{Q} \Phi Q - m \tilde{Q} Q - \mu \text{Tr}(\Phi^2).
\end{align}
We need to express the parameters $y$, $m$, and $\mu$ in the superpotential, along with the gauge coupling constant $g$ in terms of the string theory parameters, which are string scale $\ell_s \equiv \sqrt{\alpha'}$, the string coupling constant $g_s$, and the geometrical parameters of the brane configuration.
Since we aim to perform quantitative analysis of the field theory, we need to identify these parameters with correct numerical coefficients, not merely their parametric dependences. However, as we will see, our identification is consistent only when certain conditions are satisfied in addition to the weak string coupling condition $g_s \ll 1$. We discuss these conditions at the end of this section.

The distance between the two NS5-branes in the $x^6$ direction, $L_6$, controls the gauge coupling constant of the gauge theory. The gauge coupling constant $g$ is calculated from the DBI action of the D4-branes. The gauge coupling constant in the DBI action, $g_5$ is given by
\begin{align}
    \frac{1}{g_5^2} = \frac{\tau_4 (2\pi \ell_s^2)^2}{2 g_s},
\end{align}
where $\tau_p = \frac{2\pi}{(2\pi \ell_s)^{p + 1}}$ is the D$p$-brane tension in the string frame, $\ell_s$ is the string length\,\cite{Polchinski:1998rr}. The $3 + 1$ dimensional gauge coupling constant is given by compactifying the $x^6$ direction with length $L_6$:
\begin{align}\label{eq:g in terms of L6}
    \frac{1}{g^2} = \frac{L_6}{g_5^2} = \frac{L_6}{8 \pi^2 g_s \ell_s}.
\end{align}
Throughout this paper, we take the limit $g_s \ll 1$ with fixed $g$, so that the stringy effects are suppressed and the classical description of the brane configuration is valid.

Next, we consider $v'$, the distance between the D4${}_1$-brane and the D4${}_2$-brane in the $(x^4, x^5)$ plane, where we define the complex coordinate $v = x^4 + i x^5$.   
In the following discussion, we assume $g_s \ll 1$ and all the energy scales we consider are much smaller than the string scale $\ell_s^{-1}$, so that the kinetic terms of the fields on the branes are well approximated by the DBI action and assumed to be canonical.
Let us denote the positions of D4$_1$-brane and D4${}_2$-brane in the $(x^4, x^5)$ plane as $0$ and $v'$, respectively. The mass of the $W$-boson, which corresponds to the open string stretched between the two D4-branes, is given by $m_W = |v'| / (2\pi \ell_s^2)$. In the field theory picture, the mass of the $W$-boson is given by $m_W = \sqrt{2} g |\Phi^3|$. 
Here, we use the convention that the generators of the fundamental representation of $\U(2)$ are normalized as $\text{Tr}(T^a T^b) = \frac{1}{2} \delta^{ab}$ and $\Phi$ is expanded as $\Phi = \Phi^a T^a$.
Then, we have
\begin{align}
    \label{eq:phi_v'}
    \Phi = \frac{v'}{2 \sqrt{2} \pi \ell_s^2 g} \begin{pmatrix} 1 & 0 \\ 0 & 0 \end{pmatrix}.
\end{align}
In other words, we identify $\Phi^3 = \frac{v'}{2 \sqrt{2} \pi \ell_s^2 g}$ with $\Phi^0 = \Phi^3$.
Here and after, we take $x^5 = 0$ for simplicity, so that $v'$ is real.

With this identification of the adjoint scalar VEV and $v'$, we can write the correspondence between the angle of the NS5$'$-brane rotation and the adjoint mass parameter in the field theory picture. Let us define another complex coordinate $w = x^8 + i x^9$ of the $(x^8, x^9)$ plane. Before the rotation, the NS5$'$-brane is at $w' = 0$ plane, extending along the $x^1, x^2, x^3$ and $v$ directions. We rotate the NS5$'$-brane in the $(v, w)$ plane, so that the NS5$'$-brane is at 
\begin{align}
    \label{eq:NS5_rotation}
    w' = R(\phi, \theta) \, v'
\end{align}
plane with a rotation parameter $R(\theta, \phi) = e^{i\phi} \tan \theta$, where $\theta$ and $\phi$ are the rotation angles. The rotation breaks the supersymmetry down to $\mathcal N = 1$ by introducing the mass of the adjoint scalar field, $\mu$. To identify the relation between $\mu$ and $R(\theta, \phi)$, let us consider the energy density when one of the D4-branes is at $v'=0$ and the other is at $v' = \epsilon \ll \ell_s$. In the field theory picture, the potential is given by $V = |\mu|^2 |\Phi^a|^2$.
Using Eq.\,\eqref{eq:phi_v'}, we have $V = |\mu|^2 \epsilon^2 / (4 \pi^2 \ell_s^4 g^2)$, where we add both $\Phi^0$ and $\Phi^3$.
In the brane picture, the length of the D4-brane in the $x^6$ and $w$ directions is elongated by $\Delta L \simeq |R(\theta, \phi)|^2 \epsilon^2 / (2 L_6)$ for small $\epsilon$. The additional energy density is given by $\Delta E = \frac{\tau_4}{g_s} \Delta L = \frac{2\pi |R(\theta, \phi)|^2 \epsilon^2}{(2\pi \ell_s)^5 2 L_6 g_s}$. Matching the two energy densities, we have
\begin{align}
    \mu = \frac{R(\theta, \phi)}{L_6}.
\end{align}
In the following, we take $\phi = 0$ as well for simplicity, so that $R(\theta, \phi) = \tan \theta$.

\begin{figure}[t]
\begin{center}
\includegraphics[scale=0.7]{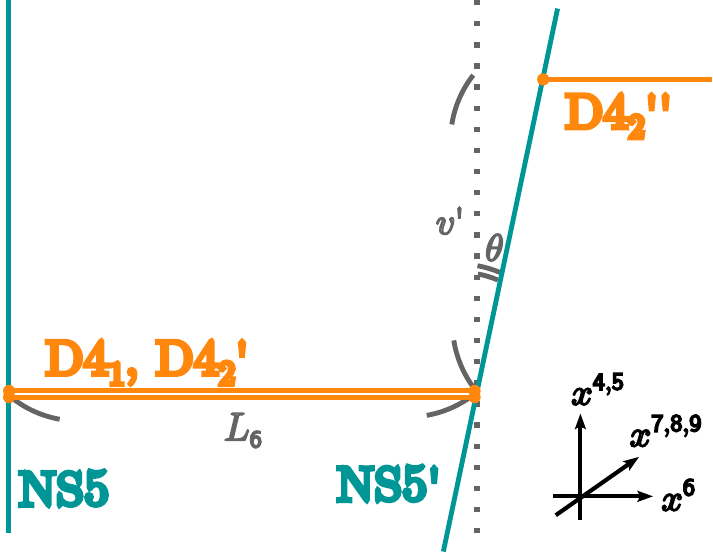}
\end{center}
\caption{
    Another vacuum of the brane configuration shown in Fig.\,\ref{fig:hanany-witten}, where D4${}_2$-brane is split into two segments, D4${}_{2}'$-brane and D4${}_{2}''$-brane. The D4${}_{2}'$-brane is hanging between the NS5-brane and the NS5$'$-brane, whereas the D4${}_{2}''$-brane is hanging between the NS5$'$-brane and extending to $x^6 \to +\infty$.
}
\label{fig:hanany-witten_broken}
\end{figure}

Next, we identify the mass parameter of the fundamental chiral multiplets.
To discuss this, let us first consider the vacua of the theory we have introduced so far.
In the brane picture shown in Fig.\,\ref{fig:hanany-witten}, the D4${}_2$-brane is semi-infinite, ending on the NS5-brane on one side and extending to $x^6 \to +\infty$. 
This brane configuration allows another vacuum, as shown in Fig.\,\ref{fig:hanany-witten_broken}.
There, the D4${}_2$-brane is split into two segments, D4${}_{2}'$-brane and D4${}_{2}''$-brane. The D4${}_{2}'$-brane is hanging between the NS5-brane and the NS5$'$-brane, being coincident with the D4${}_1$-brane so that the gauge symmetry is $\U(2)$ and unbroken. The D4${}_{2}''$-brane is hanging between the NS5$'$-brane and extending to $x^6 \to +\infty$.
One can check the total length of the D4${}_2$-brane in the $x^6$ direction is the same in both configurations, so that they are degenerate vacua of the theory. Note that one may consider D6-brane extending along the $x^1, x^2, x^3, x^7, x^8, x^9$ directions at some $x^6$ position much larger than the NS5$'$-brane position. In this case, the D4${}_2$-brane can end on the D6-brane and one can easily see that the two configurations shown in Figs.\,\ref{fig:hanany-witten} and \ref{fig:hanany-witten_broken} are indeed degenerate vacua. As long as the D6-brane is far away from the other branes, the presence of the D6-brane does not affect the dynamics of the other branes, and we can ignore it in the following discussion.

\begin{figure}[t]
\begin{center}
\includegraphics[scale=0.7]{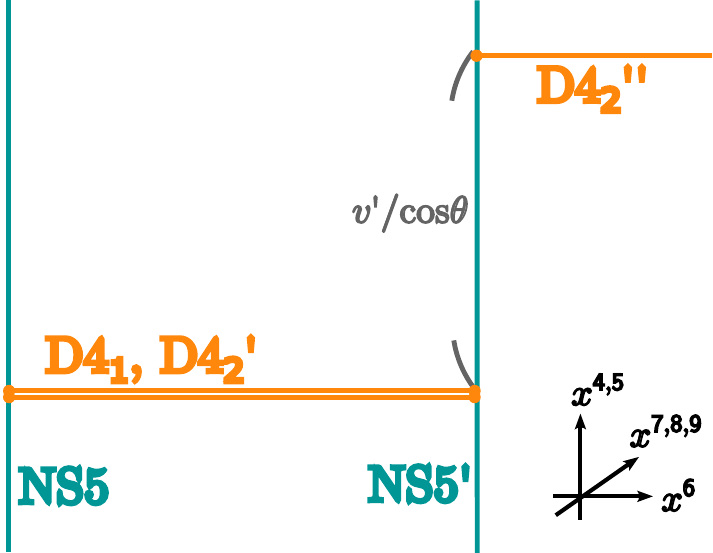}
\end{center}
\caption{
    Brane configuration realizing the $\mathcal N = 2$ supersymmetric gauge theory with $\U(2)$ gauge group with a fundamental hypermultiplet. The two NS5-branes are completely parallel, and the D4${}_2''$-brane is at $v = v' / \cos \theta$, whereas the D4${}_1$-brane and D4${}_{2}'$-brane are at $v = 0$.
}
\label{fig:hanany-witten_N=2}
\end{figure}

In the unbroken vacuum shown in Fig.\,\ref{fig:hanany-witten_broken}, the fundamental chiral multiplets corresponds to the excitations of the open strings stretched between the stack of D4-branes (D4${}_1$-brane and D4${}_{2}'$-brane) and the D4${}_{2}''$-brane. Since the two segments of the D4${}_2$-brane are at different positions in $v$ coordinate, the fundamental chiral multiplets are massive. Around the endpoint of the D4-branes on the NS5$'$-brane, the brane configuration is locally the same as that of Fig.\,\ref{fig:hanany-witten_N=2}, where the two NS5-branes are completely parallel, and the distance between the D4${}_1$-brane (and D4${}_{2}'$-brane) and the D4${}_{2}''$-brane in the $v$ coordinate is $v' / \cos \theta$. The physical mass of the hypermultiplet in this brane configuration is given by
\begin{align}
    \label{eq:hyper_mass_physical}
    m_\text{hyper}^\text{phys} = \frac{v'}{2\pi \ell_s^2 \cos\theta}.
\end{align}
Therefore, the physical mass of the fundamental chiral multiplets in the unbroken vacuum of Fig.\,\ref{fig:hanany-witten_broken} is $m_\text{hyper}^\text{phys}$ as well.

However, this is not the mass parameter $m$ appearing in the superpotential of the field theory picture, as it does not have the correct quantum number under the $R$ symmetry. In the $\mathcal N = 2$ theory shown in Fig.\,\ref{fig:hanany-witten_N=2}, the $R$-symmetry is $\SU(2)_R \times \U(1)_R$, where the $\SU(2)_R$ symmetry corresponds to the rotation in the $(x^7, x^8, x^9)$ directions and the $\U(1)_R$ symmetry does to the rotation in the $(x^4, x^5)$ directions. With the rotation of the NS5$'$-brane shown in Eq.\,\eqref{eq:NS5_rotation}, the $\SU(2)_R \times \U(1)_R$ symmetry is broken down to $\U(1)_R$ symmetry, which is the simultaneous rotation in the $v$ and $w$ planes with the same angle, $v \to e^{i\alpha} v$ and $w \to e^{i\alpha} w$.
Another $\U(1)$ symmetry is the simultaneous rotation in the $v$ and $w$ planes with $R(\theta, \phi)$, $v \to e^{i\beta} v$, $w \to e^{-i\beta} w$ and $R(\theta, \phi) \to e^{-2i\beta} R(\theta, \phi)$. Obviously, this $\U(1)$ symmetry is broken by the definite value of $R(\theta, \phi)$, but we can formally assume $R(\theta, \phi)$ is a spurion field with the $\U(1)$ charge to keep the $\U(1)$ symmetry.
Indeed, the adjoint mass in the superpotential, $W \supset - \mu \text{Tr}(\Phi^2)$, is invariant under both the $\U(1)_R$ and $\U(1)$ symmetries since $\Phi$ is the fluctuation of the D4-brane position in the $v$ direction and has the same quantum numbers as $v$.

To identify the mass parameter $m$ in the superpotential, let us first consider the $\mathcal N = 2$ limit, $\mu \to 0$. In this limit, the $\SU(2)_R \times \U(1)_R$ symmetry is restored and
the Yukawa coupling in the superpotential, $W \supset \tilde{Q} \Phi Q$, is invariant under the $\SU(2)_R \times \U(1)_R$ symmetry. Therefore, the Yukawa coupling is also invariant under both the $\U(1)_R$ and $\U(1)$ symmetries. The mass term of the fundamental chiral multiplets in the superpotential, $W \supset - m \tilde{Q} Q$, should also be invariant under both the $\U(1)_R$ and $\U(1)$ symmetries; in other words, the mass parameter $m$ should have the same quantum numbers as $\Phi$, or $v$.
Therefore, the mass term such as Eq.\,\eqref{eq:hyper_mass_physical}, or $v' / R(\theta, \phi)$, to maintain the holomorphicity, cannot enter the superpotential as it has different quantum numbers under the $\U(1)$ symmetries. Instead, the mass parameter $m$ in the superpotential should be proportional to $v'$, which has the same quantum numbers as $v$.
Matching the dimensions, we have
\begin{align}
    m = \frac{v'}{2\pi \ell_s^2}.
\end{align}
Then, to have the correct physical mass, we assume that the K\"ahler potential of the fundamental chiral multiplets, $\mathcal{K}_Q$, is non-trivial and given by
\begin{align}
    \mathcal{K}_Q = \cos\theta (Q^\dagger e^V Q + \tilde{Q}^\dagger e^{-V} \tilde{Q}),
    \label{eq:kahler_potential_Q}
\end{align}
where $V$ is the vector superfield. With this K\"ahler potential, the physical mass of the fundamental chiral multiplets is given by $m_\text{hyper}^\text{phys} = |m / \cos \theta|$, which matches Eq.\,\eqref{eq:hyper_mass_physical}.

Finally, let us identify the Yukawa coupling constant between the adjoint chiral multiplet and the fundamental chiral multiplets.
Let the Yukawa coupling constant be $y$. With the full superpotential shown in Eq.\,\eqref{eq:superpotential_full}, the $F$-flatness condition for $\Phi, Q$ and $\tilde{Q}$ requires
\begin{gather}
    \frac{y}{\cos\theta} Q \tilde{Q} - 2 \mu \Phi = 0, \quad
    y \Phi Q - m Q = 0, \quad
    y \tilde{Q} \Phi - m \tilde{Q} = 0.
\end{gather}
Together with the $D$-flatness condition, we can fix $y$ and the VEVs of the chiral multiplets:
\begin{gather}
    y = \sqrt{2} g, \quad
    Q = \begin{pmatrix} q \\ 0 \end{pmatrix}, \quad
    \tilde{Q} = \begin{pmatrix} q & 0 \end{pmatrix},
\end{gather}
with
\begin{align}
    q^2 = \frac{v'\mu \cos\theta}{2 \pi \ell_s^2 g^2}.
\end{align}

Let us then discuss the conditions for the validity of the above identification of the parameters.
First, we need to require that the string coupling constant is weak, so that the stringy effects are suppressed and the classical description of the brane configuration is valid.
Since we would like to fix the gauge coupling constant $g$ in the field theory picture, we need to take the limit
\begin{align}
    \label{eq:weak_string_coupling}
    g_s \to 0, \quad L_6 / \ell_s \to 0, 
\end{align}
simultaneously, with fixed $g$. In the numerical analysis below, we take finite values, $g_s = 10^{-3}$ and $L_6/\ell_s = 0.3$ for the reference parameters (up to $0.6$ in the scaling checks of Sec.\,\ref{sec:numerical_calculation}), which give $g \simeq 0.5$; $L_6/\ell_s$ is thus small but not parametrically small. 
Next, with the above identification of the parameters, we first require that the physical masses of the fields on the branes in the unbroken vacuum are much smaller than heavier modes such as the stringy modes and the Kaluza-Klein modes in the $x^6$ direction, so that the low-energy effective field theory on the branes is well approximated by the field theory picture we have introduced. This requires
\begin{align}
    \label{eq:low_energy_effective_field_theory}
    \tan \theta \ll \pi, \quad \frac{v'}{2\pi \ell_s} \ll 1,
\end{align}
given that we take the limit $L_6 / \ell_s \to 0$.

With the above conditions, both the brane picture and the field theory picture are valid descriptions of the same physics. 
Since we are interested in the metastable vortex strings,
the scale of the first symmetry breaking $\U(2) \to \U(1)\times \U(1)$ should be larger than that of the second symmetry breaking $\U(1)\times \U(1) \to \U(1)$. Therefore, we need to require that
\begin{align}
    \qty(\frac{v'}{2 \sqrt{2} \pi \ell_s^2 g})^2 > \qty(\frac{q}{2})^2,
\end{align}
where the factor of $1/2$ comes from the normalization of the generators of the fundamental representation of $\SU(2)$.
Equivalently, we have
\begin{align}
    \label{eq:meta-stable_vortex_condition}
    \frac{v'}{\pi \ell_s^2} \gg \mu.
\end{align}
With this condition, the metastable vortex strings appear in the field theory picture. 

In the field theory picture, metastable vortex strings and monopoles appear as the result of the hierarchical symmetry breaking, $\U(2) \to \U(1)\times \U(1) \to \U(1)$.
The mass of the monopole and the tension of
the vortex string are\,\cite{Shifman:2012zz} given by
\begin{align}
    m_\text{monopole} \sim \frac{2v'}{g^2\ell_s^2}
    ,\quad
    T_\text{vortex} \sim 4\pi q^2,
    \label{eq:mass_topological_defects}
\end{align}
respectively, up to non-BPS corrections.

Since both the brane picture and the field theory picture are valid descriptions of the same physics, the metastable vortex strings and the monopoles should appear in the brane picture as well. We will discuss the monopole in the brane picture in the next section, and here we focus on the metastable vortex strings.
Unlike the BPS vortex strings in the $\mathcal N = 2$ supersymmetric gauge theories\,\cite{Hanany:2003hp,Hanany:2004ea}, the metastable vortex strings are non-BPS, so that they do not correspond to BPS configuration of D-branes in the brane picture. However, based on the observation in Ref.\,\cite{Auzzi:2004yg}, we can identify the metastable vortex strings with the non-BPS configuration of D2-branes
stretched between the NS5$'$-brane and the D4${}_2$-brane and infinite in the, for example, $x^3$ direction, as shown in Fig.\,\ref{fig:hanany-witten}. Then, the tension of the metastable vortex string in the brane picture is 
\begin{align}
    T_3 = \frac{\tau_2}{g_s} \times v' \sin \theta \simeq 4 \pi q^2. 
    \label{eq:tension_brane}
\end{align}
Here, however, we have ignored the bending of the D4${}_2$-brane due to the presence of the D2-brane. As $L_6 < \ell_s$, the bending is not negligible\,\cite{Callan:1997kz,Gibbons:1997xz}, and the tension of the D2-brane is smaller than that in Eq.\,\eqref{eq:tension_brane}. This bending is the brane counterpart of the classical profile of the fields around the vortex string, just as a D-string ending on a D3-brane bends it into the profile of the BPS monopole\,\cite{Callan:1997kz,Gibbons:1997xz}; neglecting it amounts to neglecting the core structure of the vortex string. The bending is not controlled by the conditions in Eqs.\,\eqref{eq:low_energy_effective_field_theory} and \eqref{eq:meta-stable_vortex_condition}, and we come back to its possible effects in Sec.\,\ref{sec:conclusion_and_discussion}. 

Now, let us compare the tension of the metastable vortex in the brane picture with that in the field theory picture. 
The order of magnitude of the tension matches, as in Ref.\,\cite{Auzzi:2004yg}, where the tension is identified up to a numerical factor. The question is whether the numerical coefficient matches as well. To check this, we explicitly construct the string solution in the field theory picture and calculate the tension of the vortex string. The details of the construction of the vortex string solution and the calculation of its tension are shown in Appendix\,\ref{sec:vortex_solution}. In Fig.\,\ref{fig:tension}, we show the ratio of the tension of the vortex string in the field theory picture, $T_\text{vortex}$, to that in the brane picture, $T_3$, as a function of $\theta$. Here, we take $L_6 = 6$, $v' = 25$, $\ell_s = 20$ and $g_s=10^{-3}$, which satisfy Eqs.\,\eqref{eq:low_energy_effective_field_theory} and \eqref{eq:meta-stable_vortex_condition}. We can see that the ratio approaches $1$ as $\theta \to 0$ and stays within about $15\%$ for $\theta \leq 0.1$, although we have ignored the bending of the D4${}_2$-brane in the brane picture. This
is a non-trivial check of the identification of the metastable vortex string with the non-BPS D2-brane in the brane picture. 
The ratio is in fact bounded from above. Setting $\eta = 1$ in the tension functional of Appendix\,\ref{sec:vortex_solution} freezes the adjoint field at its vacuum value, and the remaining Abelian-Higgs model has the BPS potential, whose vortex tension is $4\pi q^2 = T_3$ up to $\mathcal O(\theta^2)$ corrections from the K\"ahler potential. Since the actual solution minimizes the functional, $T_\text{vortex} \leq T_3$. The deviation at finite $\theta$ comes from the relaxation of the adjoint field inside the core of the vortex string, $\eta < 1$, which is suppressed as $\theta \to 0$, where the adjoint field becomes much heavier than the fields of the second symmetry breaking. In the brane picture, this relaxation is a displacement of the D4${}_2$-brane in the $v$ direction near the D2-brane. Heuristically, the D2-brane is perpendicular to the NS5$'$-brane and pulls the D4${}_2$-brane with a component along $v$ proportional to $\sin\theta$, which vanishes as $\theta \to 0$. This part of the bending, neglected in Eq.\,\eqref{eq:tension_brane}, lowers the tension, consistent with the sign of the deviation.

\begin{figure}[t]
\begin{center}
\includegraphics[scale=0.7]{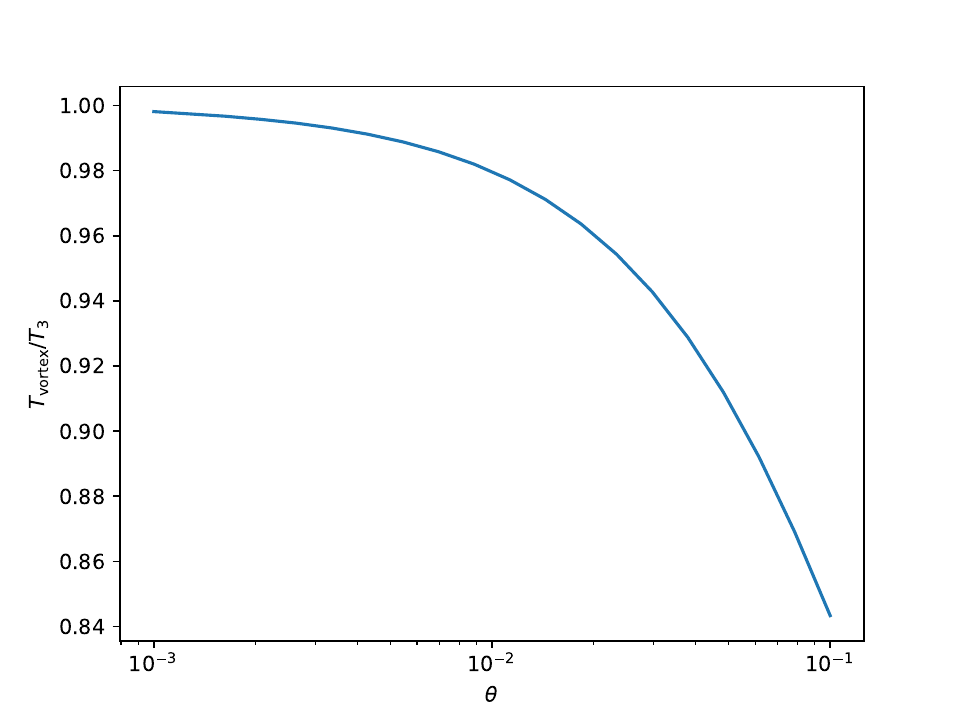}
\end{center}
\caption{
    The ratio of the tension of the vortex string in the field theory picture, $T_\text{vortex}$, to that in the brane picture, $T_3$, as a function of $\theta$. Here, we take $L_6 = 6$, $v' = 25$, $\ell_s = 20$ and $g_s=10^{-3}$, which satisfy Eqs.\,\eqref{eq:low_energy_effective_field_theory} and \eqref{eq:meta-stable_vortex_condition}.
}
\label{fig:tension}
\end{figure}

\section{String breaking processes in the brane picture}
\label{sec:string_breaking}
With the setup we have introduced in the previous section, we can analyze the string breaking processes in the brane picture. In the field theory picture, the metastable vortex string can break by creating a monopole-anti-monopole pair; the homotopy class of the second symmetry breaking $\U(1)\times \U(1) \to \U(1)$ is non-trivial, $\pi_1(\U(1)\times \U(1) / \U(1)) = \mathbb{Z}$, but the homotopy class of the total symmetry breaking $\U(2) \to \U(1)$ is trivial, $\pi_1(\U(2) / \U(1)) = 0$. Correspondingly, in the brane picture, the non-BPS D2-brane can shrink to nothing.
Then, what is the string breaking process in the brane picture? 

\begin{figure}[t]
\centering
\begin{minipage}{0.42\textwidth}
\includegraphics[scale=0.55]{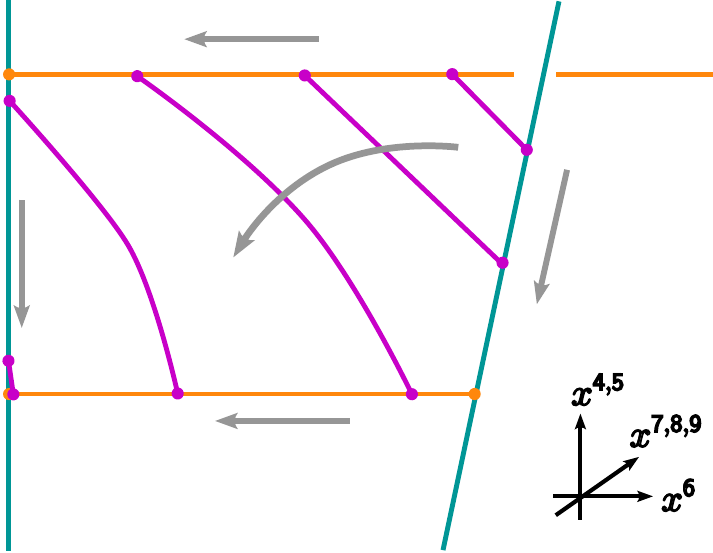}
\end{minipage}
\quad \quad
\begin{minipage}{0.42\textwidth}
\includegraphics[scale=0.55]{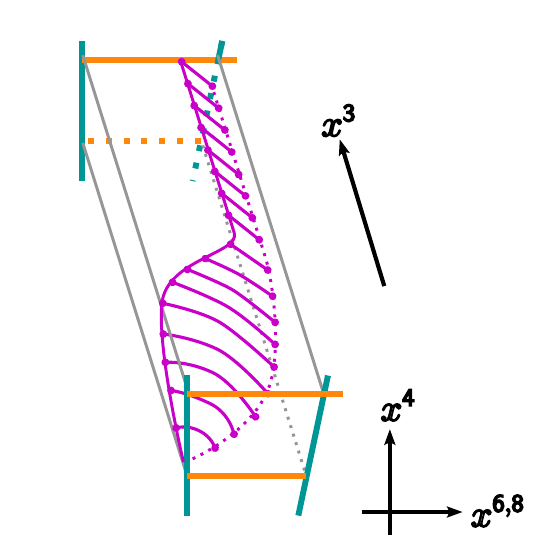}
\end{minipage}
\caption{
    A schematic picture of the string end in the brane picture. The configuration of the D4-branes and the NS5-branes is the same as that in Fig.\,\ref{fig:hanany-witten}. The D2-brane, which is shown in magenta, first hangs between the NS5$'$-brane and the D4${}_2$-brane, which may shrink to nothing as we go along the string direction to the monopole end.
    In the left figure, we show the configuration of the branes in the $x^{4, \cdots , 9}$ directions, where the D2-brane is shown in magenta.
    The gray arrows show the path along the direction of the string.
    In the right figure, we project the brane configuration in the $(x^4, x^{6,8})$ plane and show the $x^3$ direction as well.
}
\label{fig:hanany-witten_string_end}
\end{figure}

First, we show the schematic picture of the end of the broken string straight in the $x^3$ direction in the brane picture in Fig.\,\ref{fig:hanany-witten_string_end}. 
The D2-brane first hangs between the NS5$'$-brane and the D4${}_2$-brane, which corresponds to the metastable vortex string. However, the energy of the D2-brane can be as small as we want; the smallest energy configuration is when the D2-brane shrinks to nothing. The path from the D2-brane configuration to the configuration with no D2-brane is shown by the gray arrows in the left figure, which corresponds to the path along the direction of the string to its end, as shown in the right figure. 
The length of the D2-brane first increases as the path goes from the initial configuration, which corresponds to the energy of the string end, the monopole\,\cite{Auzzi:2004yg,Eto:2012kb}. 
With this identification of the monopole in the brane picture, we may calculate the mass of the monopole in the brane picture. Obviously, it is impossible to separate the monopole end and the string itself, since they are continuously connected, but we can approximately regard the monopole as the configuration of the D2-brane apart from the vortex string, the D2-brane stretched between the NS5${}'$-brane and the D4${}_2$-brane. Then, when $\theta$ is small enough, the mass of the monopole is approximately given by the energy of the D2-brane stretched between the NS5-branes and the D4${}_2$-brane, $m_\text{monopole} \simeq \tau_2 v' L_6 / g_s$. This indeed matches with the mass of the monopole in the field theory picture in Eq.\,\eqref{eq:mass_topological_defects}.

\begin{figure}[t]
\centering
\begin{minipage}{0.45\textwidth}
    \centering
    \includegraphics[scale=0.15]{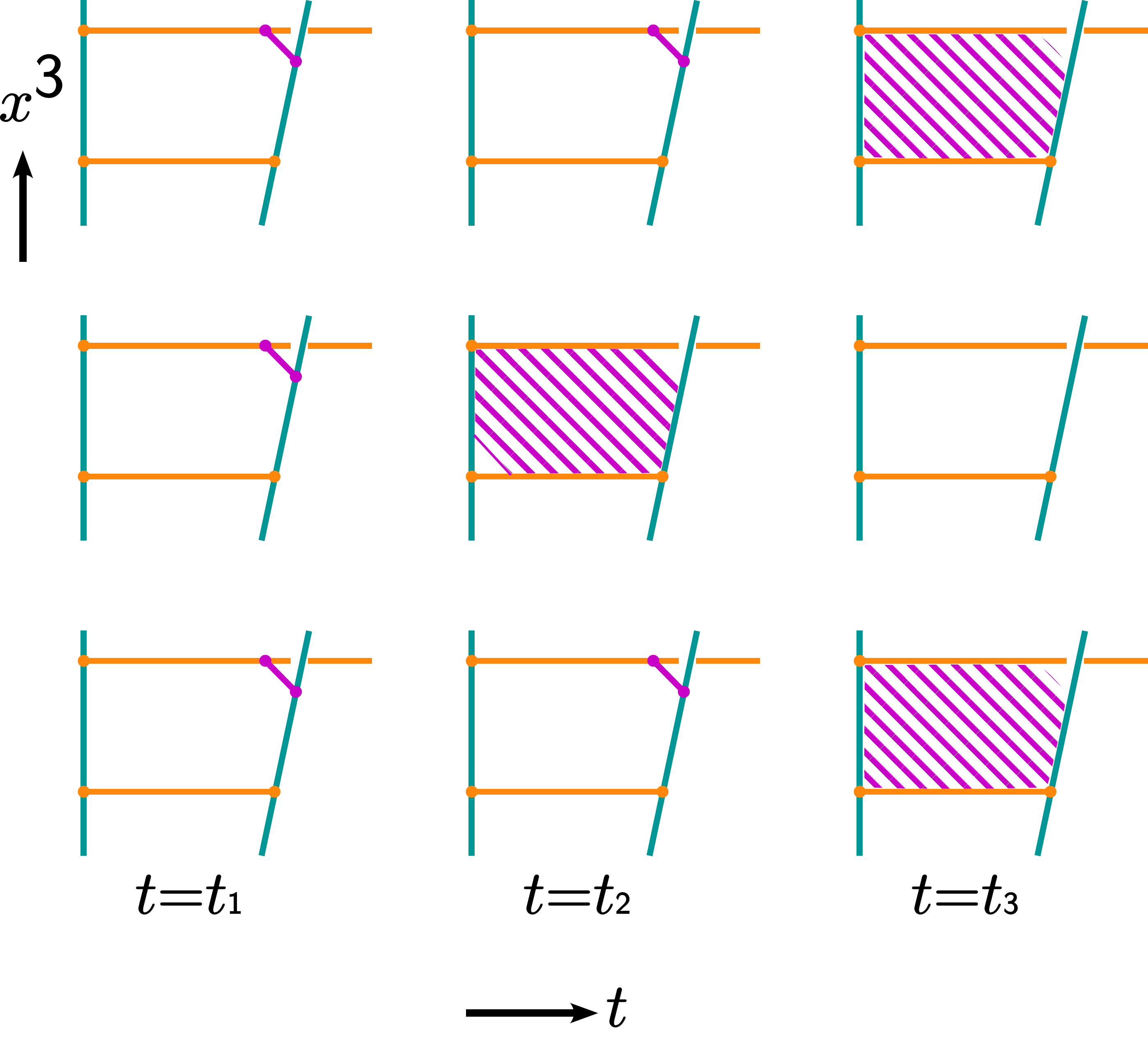}
\end{minipage}
\quad \quad \quad
\begin{minipage}{0.45\textwidth}
    \centering
    \includegraphics[scale=0.15]{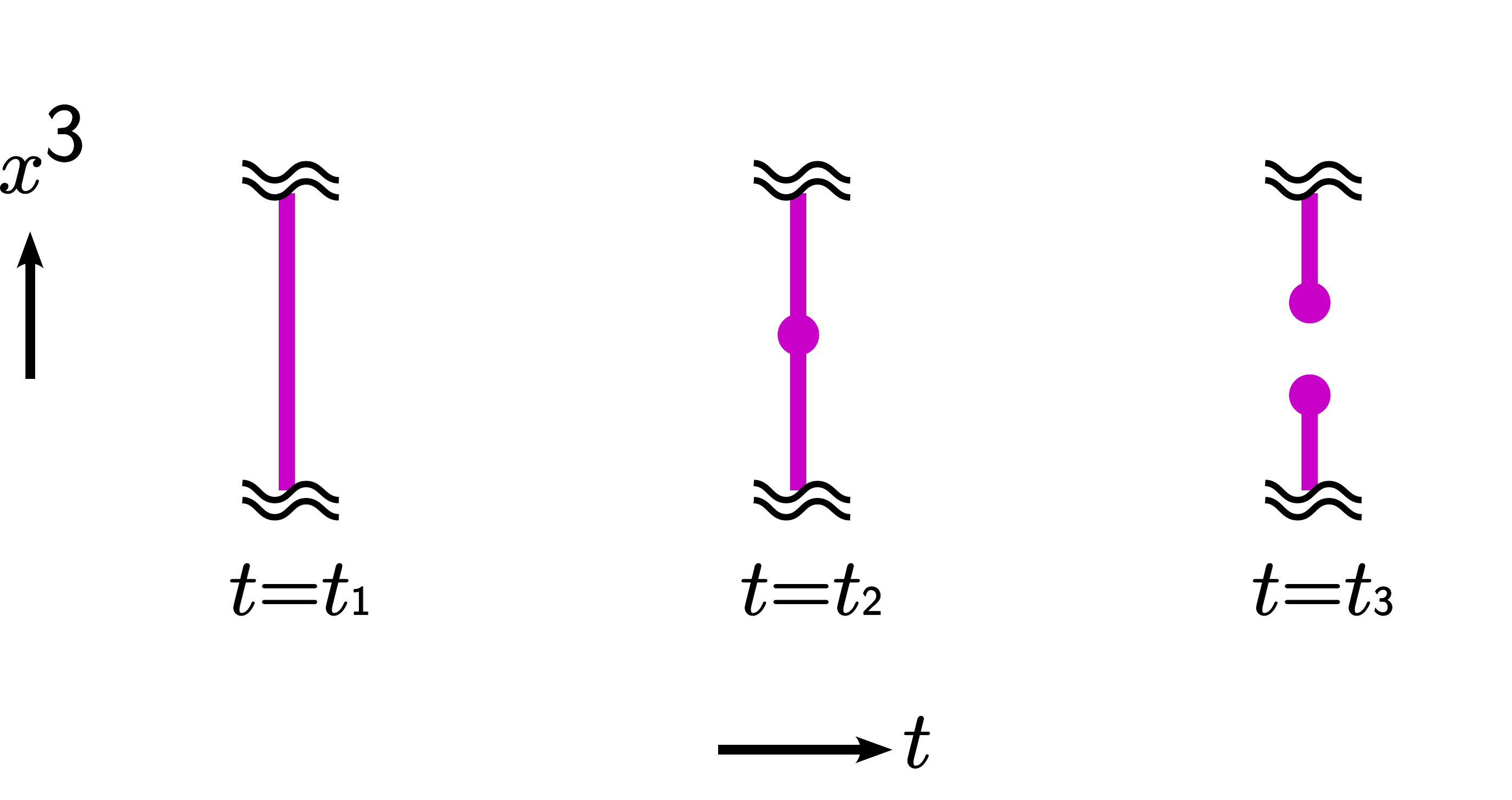}
\end{minipage}
\caption{
    A schematic picture of the tunneling process of the string breaking in the brane picture (left) and in the field theory picture (right).
    In the brane picture, we show the configuration of the D4-branes and the NS5-branes, which is the same as that in Fig.\,\ref{fig:hanany-witten}, together with the D2-brane, which is shown in magenta, for some points in the $(t, x^3)$-space.
    The corresponding configuration of the vortex string in the field theory picture is shown on the right, where the vortex string is shown in magenta.
    In both pictures, at $t=t_2$, the instanton process occurs, and the string breaks by creating a monopole-anti-monopole pair. In the right picture, both the instanton process and the monopole-anti-monopole pair are shown by the magenta blobs.}
\label{fig:tunneling}
\end{figure}

With this picture of the string end, the string breaking process can be understood as the tunneling process of the D2-brane from the initial configuration to the configuration with no D2-brane. As an example, we show the schematic figure of the breaking process of a straight string in $x^3$ direction in Fig.\,\ref{fig:tunneling}. In the left figure, we show the configuration of the vortex string in the brane picture whereas in the right figure, we show the corresponding configuration of the vortex string in the field theory picture. 
We take some points in the $(t, x^3)$-space and show the configuration of the branes at those points. At $t = t_1$, the D2-brane is hanging between the NS5$'$-brane and the D4${}_2$-brane for any $x^3$ position, which corresponds to the straight vortex string. At $t = t_2$, the D2-brane shrinks to nothing at some $x^3$ position in the brane picture, which corresponds to the instanton process in the field theory picture. For the D2-brane to shrink to nothing, we need to tunnel through the energy barrier, so that the process in the brane picture is also the instanton process. After the tunneling process, at $t = t_3$, the straight string is broken into two pieces, and the monopole-anti-monopole pair is created at the end of both pieces.

Before discussing how to calculate the bounce action of the metastable vortex string in the brane picture, let us first mention analytical estimates of the decay rate in terms of $m_\text{monopole}$ and $T_\text{vortex}$. The decay rate of the metastable vortex string is given by $\Gamma \sim e^{-S_\text{inst}}$ with $S_\text{inst}$ being the action of the instanton corresponding to the tunneling process, which we also call the bounce action. The action of the instanton can be analytically estimated by using the thin-wall approximation\,\cite{Vilenkin:1982hm,Preskill:1992ck}. In the thin-wall approximation, we approximate the total action of the system as the sum of the action of the string and the action of the monopole-anti-monopole pair:
\begin{align}
    S = m_\text{monopole}     \int ds + T_\text{vortex}     \int dA,
\end{align}
where $ds$ is the line element along the worldline of the monopole, and $dA$ is the area element of the worldsheet of the string.
For the straight string in the $x^3$ direction, we assume $\text{SO}(2)$ symmetry in the $(t, x^3)$-space. Then, the tunneling action is given as a function of the radius of the ``bubble'' on the string worldsheet, $R$, as
\begin{align}
    \label{eq:thin_wall_ft}
    S_\text{inst}(R) = 2\pi m_\text{monopole} R - \pi T_\text{vortex} R^2.
\end{align}
Here, we subtract the action of the metastable string. By minimizing $S_\text{inst}(R)$ with respect to $R$, we have
\begin{align}
    \label{eq:thin_wall_ft_result}
    R_\text{inst} = \frac{m_\text{monopole}}{T_\text{vortex}}, \quad S_\text{inst} = \pi \frac{m_\text{monopole}^2}{T_\text{vortex}}.
\end{align}

Then, the next question is how to calculate $S_\text{inst}$ in the brane picture.
In the brane picture, $S_\text{inst}$ is given by the action of the instanton corresponding to the tunneling process of the D2-brane. 
The action of the instanton should be calculated from the DBI action of the D2-brane if the motion of the D2-brane corresponding to the tunneling process is given. The DBI action of the D2-brane is given by
\begin{align}
    \label{eq:DBI_action}
    S_\text{DBI} = \frac{\tau_2}{g_s} \int d^3 \xi \sqrt{\det \partial_\alpha X^\mu \partial_\beta X_\mu},
\end{align}
where $\xi^\alpha$ are the worldvolume coordinates of the D2-brane and $X^\mu$ are the spacetime coordinates. We have ignored the contribution from the gauge field on the D2-brane and the NS-NS $B$-field. We also assume that the spacetime metric is flat and ignore the backreaction of the branes. Note that the DBI action has a gauge symmetry corresponding to the diffeomorphism of the worldvolume coordinates, so that we can choose any gauge to calculate the action. We will take the static gauge later.
We then need to find the tunneling path of the D2-brane, whose ends are fixed on the NS5-branes and the D4-branes shown in Fig.\,\ref{fig:tunneling}, and calculate the DBI action along the tunneling path.

\begin{figure}[t]
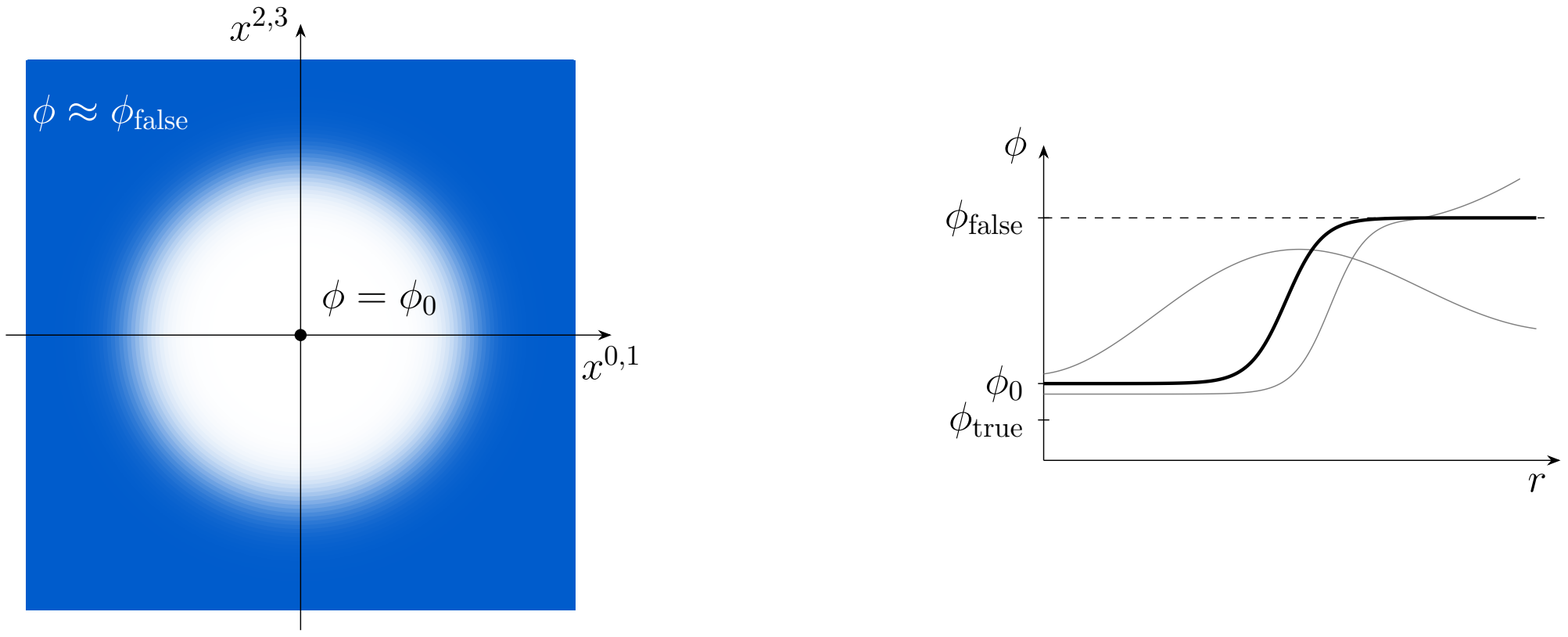

\centering
\includestandalone[width=0.8\textwidth]{tikz/fig_scalar_bounce}
\caption{%
    A schematic picture of the bounce solution for the false vacuum decay of a
    scalar field $\phi$. In the left figure, we show the $\text{O}(4)$-symmetric
    bounce configuration on the Euclidean spacetime. The field takes the value
    $\phi = \phi_0$, close to the true vacuum, at the center (black dot), and
    approaches the false vacuum, $\phi \approx \phi_\text{false}$, far from the
    center, through the wall region where the field varies rapidly. In the right
    figure, we show the profile of the bounce solution (thick curve), which
    starts at $\phi_0$ with $\partial_r \phi = 0$ and asymptotes to
    $\phi_\text{false}$ (dashed line). The thin curves show solutions with
    slightly different initial values, which overshoot or undershoot the false
    vacuum.%
}
\label{fig:bounce_scalar}
\end{figure}

To see how to find the tunneling path, let us first review the calculation of the decay rate of the false vacuum in quantum field theory\,\cite{Coleman:1977py,Callan:1977pt}. Suppose we have a scalar field $\phi$ with a potential $V(\phi)$, which has a local minimum at $\phi = \phi_\text{false}$ and a global minimum at $\phi = \phi_\text{true}$. The instanton corresponding to the tunneling process from the false vacuum to the true vacuum is given by the bounce solution $\phi_\text{bounce}$, which is a solution of the Euclidean equation of motion with appropriate boundary conditions, $\phi_\text{bounce} \to \phi_\text{false}$ at $|x| \to \infty$ and $\partial_\mu \phi_\text{bounce} = 0$ at $x = 0$. Here, we have assumed that the bounce solution is $\text{O}(4)$ symmetric, so that the solution depends only on the radial coordinate $r = \sqrt{x_\mu x^\mu}$. The decay rate of the false vacuum is given by $\Gamma \sim e^{-S_\text{bounce}}$ with $S_\text{bounce}$ being the Euclidean action of the bounce solution. The simplest way to find the bounce solution is so-called overshoot/undershoot method. We first choose some initial value of the field at $r = 0$, $\phi_\text{bounce}(0) = \phi_0$, and solve the Euclidean equation of motion with the boundary condition $\partial_\mu \phi_\text{bounce} = 0$ at $r = 0$. If $\phi_0$ is sufficiently close to $\phi_\text{true}$, the solution overshoots the false vacuum at some finite $r$, whereas if $\phi_0$ is sufficiently far from $\phi_\text{true}$, the solution undershoots the false vacuum and starts to oscillate at some finite $r$. By tuning $\phi_0$, we can find the bounce solution, which approaches the false vacuum at $r \to \infty$. 
We schematically depict this solution in Fig.\ \ref{fig:bounce_scalar}.

We try to apply these methods to find the tunneling path of the D2-brane. First, we perform the Wick rotation to the DBI action, $t \to -i \tau$, to rewrite the action in the Euclidean signature. Then, just like the $\text{O}(4)$ symmetry in the bounce solution, we may assume that the tunneling path of the D2-brane corresponding to the straight vortex string in the $x^3$ direction is $\text{O}(2)$ symmetric in the $(\tau, x^3)$-space. We rewrite the coordinates as $\tau = \rho \cos \alpha$ and $x^3 = \rho \sin \alpha$ with $\rho = \sqrt{\tau^2 + (x^3)^2}$ and $\alpha = \arctan(x^3 / \tau)$, and assume that the tunneling path of the D2-brane is independent of $\alpha$. We next take the static gauge, $\xi^0 = \tau$ and $\xi^1 = x^3$. The final coordinate, $\xi^2\equiv s$, is defined as the coordinate along the direction of the string; without loss of generality, we can take $s \in [0, 1]$ and $s = 0$ and $s = 1$ correspond to the ends of the D2-brane. With these simplifications, the DBI action is reduced to
\begin{align}
    \label{eq:DBI_action_reduced}
    S_\text{DBI} = \frac{2\pi \tau_2}{g_s} \bigintss_0^\infty d\rho \bigintss_0^1 ds \rho \sqrt{\qty[1 + \qty(\frac{\partial \vec{X}}{\partial \rho})^2] \qty(\frac{\partial \vec{X}}{\partial s})^2 - \qty(\frac{\partial \vec{X}}{\partial \rho} \cdot \frac{\partial \vec{X}}{\partial s})^2},
\end{align}
where $\vec{X}(\rho, s)$ are the coordinates of the D2-brane in the directions transverse to the $(\tau, x^3)$-space. 

One may then try to find the bounce solution of the DBI action, which approaches the configuration of the D2-brane corresponding to the straight vortex string at $\rho \to \infty$, starting from some initial configuration sufficiently close to the true vacuum at $\rho = 0$.
However, this does not work, as the true vacuum configuration corresponds to the configuration with no D2-brane and cannot be described by the DBI action of the D2-brane.
The true stable vacuum should consist of some stringy excitations in the brane background. Therefore, instead of changing the initial field value at $r = 0$ in the vacuum decay calculation, \emph{we claim that the bounce solution suddenly starts to exist at some finite $\rho = \rho_0$, with a short enough D2-brane}. The change of the initial field value at $r = 0$ corresponds to
not only the change of the initial configuration of the D2-brane such as its velocity and its shape, but also the change of $\rho_0$, the time when the D2-brane suddenly appears (See Fig.\ \ref{fig:bounce_brane} left panel). The overshoot/undershoot method in the brane picture is then to tune the initial configuration of the D2-brane and $\rho_0$ to find the solution of the Euclidean equation of motion of the DBI action, which approaches the configuration of the D2-brane corresponding to the straight vortex string at $\rho \to \infty$. 

Before constructing the bounce solution explicitly, we derive its action within the thin-wall approximation in this brane framework, by assuming that in the bounce solution the variation of D2-brane configuration sharply concentrates around $\rho \sim \rho_0$ for some $\rho_0$. This means that the bounce action behaves as
\begin{equation}
    \label{eq:thin_wall_brane}
    S_\text{DBI} \approx \frac{\tau_2}{g_s} \left(2\pi\rho_0 A_\text{sweep} - \pi \rho_0^2 v' \sin\theta\right),
\end{equation}
where $A_\text{sweep}$ is the area of the surface the D2 brane sweeps in the space spanned by $x^{4,5,6,7,8,9}$ coordinates during the decay. The first term is the worldvolume swept by the D2-brane along the circle $\rho = \rho_0$ in the $(\tau, x^3)$-space, whose circumference is $2\pi \rho_0$. The second term is from the loss of action due to the absence of the D2-brane in the region $\rho \le \rho_0$, whose area is $\pi \rho_0^2$, the D2-brane there being of width $v' \sin \theta$.
While the area $A_\text{sweep}$ should be obtained by minimizing the area among the potential trajectory of the D2 brane (in Ref.\,\cite{Eto:2012kb}, such minimal surfaces give the monopole masses in a different brane configuration),
when $\theta$ is small we can approximate the minimal area surface by the rectangle formed by NS5 and D4 branes projected onto $x^{4,5,6}$ directions, whose area is $L_6 v'$.
Thus in this approximation we have
\begin{equation}
    \label{eq:thin_wall_brane_small_theta}
    S_\text{DBI}(\rho_0)\approx \frac{\tau_2}{g_s}\left(2\pi \rho_0 L_6 v' - \pi \rho_0^2 v'\sin\theta\right).
\end{equation}

Comparing Eq.\,\eqref{eq:thin_wall_brane} with the field theory result of Eq.\,\eqref{eq:thin_wall_ft}, we see that the two expressions agree term by term, once we identify
\begin{align}
    \label{eq:thin_wall_identification}
    m_\text{monopole} = \frac{\tau_2}{g_s} A_\text{sweep}, \quad T_\text{vortex} = \frac{\tau_2}{g_s} v' \sin\theta = T_3.
\end{align}
The identification of the string tension is nothing but Eq.\,\eqref{eq:tension_brane}, which we have already checked against the explicit vortex string solution in Fig.\,\ref{fig:tension}. As for the monopole mass, using the small $\theta$ approximation $A_\text{sweep} \simeq L_6 v'$ together with Eq.\,\eqref{eq:g in terms of L6}, we obtain
\begin{align}
    m_\text{monopole} \simeq \frac{\tau_2 v' L_6}{g_s} = \frac{2 v'}{g^2 \ell_s^2} = \frac{4 \pi m_W}{g^2},
\end{align}
which is the mass of the monopole in the field theory picture quoted in Eq.\,\eqref{eq:mass_topological_defects}. In other words, the thin-wall approximation in the brane picture reproduces the field theory one of Eqs.\,\eqref{eq:thin_wall_ft} and \eqref{eq:thin_wall_ft_result}.

The radius of the bubble and the bounce action in this approximation are then given by
\begin{align}
    \label{eq:thin_wall_brane_result}
    \rho_0 = \frac{A_\text{sweep}}{v' \sin\theta} \simeq \frac{L_6}{\sin\theta}, \quad
    S_\text{inst} \simeq \pi \frac{\tau_2}{g_s} \frac{L_6^2 v'}{\sin\theta} = \frac{2\pi v' L_6}{g^2 \ell_s^2 \sin\theta}.
\end{align}
Note that the string scale $\ell_s$ and the string coupling constant $g_s$ enter these expressions only through the gauge coupling constant $g$. Indeed, using $\mu = \tan\theta / L_6$ and taking $\theta$ small, we have $\rho_0 \simeq 1 / \mu$ and $S_\text{inst} \simeq 2 \pi v' / (g^2 \ell_s^2 \mu)$, which are written purely in terms of the field theory quantities. This is the analytic counterpart of the invariance of the bounce action under the scalings of the brane parameters, which we check numerically in Sec.\,\ref{sec:numerical_calculation}. The validity of the thin-wall approximation itself is discussed in Sec.\,\ref{sec:numerical_calculation}, where the numerical solutions let us estimate the thickness of the ``wall''.

\section{Numerical calculation and results}
\label{sec:numerical_calculation}
As a consistency check of the thin-wall result, we now construct the bounce solution numerically, based on the conjecture of the bounce solution in the previous section. To simplify the calculation, we make one key assumption: the D2-brane remains straight in the space of $x^{4,5,6,7,8,9}$, with its shape entirely determined by the positions of its endpoints.
In principle, the D2-brane can be curved, and its shape is determined by solving the equation of motion from the DBI action. However, we expect this straight-brane approximation to be valid in the limit of small $\theta$.
The endpoints of the D2-brane are fixed on the NS5-branes and the D4-branes, so that they can be described by one-parameter functions: $\vec{X} = \vec{X}_E(u)$. We normalize the parameter $u$ such that $d\vec{X}_E/du$ is a unit vector along the branes, with $\vec{X}_E(0)$ at the intersection of the NS5-brane and D4${}_1$-brane, $\vec{X}_E(v')$ at the intersection of the NS5-brane and D4${}_2$-brane, $\vec{X}_E(-L_6)$ at the intersection of the NS5$'$-brane and D4${}_1$-brane, and so forth. The kinematic degrees of freedom of the D2-brane are then the endpoint parameters as functions of $\rho$, which we denote as $u_0(\rho)$ and $-u_1(\rho)$, with both $u_0$ and $u_1$ non-negative at the initial configuration. For given $(\rho, s)$, the field $\vec{X}(\rho, s)$ is given by the linear interpolation of the endpoints, $\vec{X}(\rho, s) = (1 - s) \vec{X}_E(u_0(\rho)) + s \vec{X}_E(-u_1(\rho))$.
We note that this parameterization of the D2-brane is non-differentiable at the edges, the intersection of a NS5-brane and a D4-brane. In the numerical calculation, we smooth the edge by a small smoothing parameter $\epsilon$; instead of rectangular corners, we use quarter circles with radius $\epsilon$ at the edges. $\epsilon$ is taken to be much smaller than any other length scale in the problem, and we check that the results do not depend on $\epsilon$.

With this straight-brane approximation, the DBI action is reduced to a functional of $u_0(\rho)$ and $u_1(\rho)$ as
\begin{align}
    S_\text{DBI} = \frac{2\pi \tau_2}{g_s} \int_0^\infty d\rho \mathcal{L}, \quad \mathcal{L} = \rho L(u_0, u_1) I(u_0, u_1, \dot{u}_0, \dot{u}_1),
\end{align}
where
\begin{align}
    \vec{L} &\equiv \vec{X}_E(-u_1) - \vec{X}_E(u_0), \nonumber \\
    I &\equiv \int_0^1 ds \sqrt{1 + M^2 - (\vec{\hat{L}} \cdot \vec{M})^2}, \nonumber \\ 
    \vec{M}  &\equiv \frac{\partial \vec{X}}{\partial \rho} = 
    \dot{u}_0 (1 - s) \vec{v}_0 + \dot{u}_1 s \vec{v}_1 \nonumber \\ 
    \vec{v}_0 &\equiv \frac{\partial \vec{X}_E(u_0)}{\partial u_0}, \quad
    \vec{v}_1 \equiv \frac{\partial \vec{X}_E(-u_1)}{\partial u_1},
\end{align}
with the dot denoting the derivative with respect to $\rho$, $\vec{\hat{A}} \equiv \vec{A} / A$ and $A = |\vec{A}|$ for any vector $\vec{A}$. The equations of motion for $u_0(\rho)$ and $u_1(\rho)$ are given by
\begin{align}
    \label{eq:eom_for_u}
    \mathcal{I} \begin{pmatrix} \ddot{u}_0 \\ \ddot{u}_1 \end{pmatrix} = \begin{pmatrix} \mathcal{F}_0 \\ \mathcal{F}_1 \end{pmatrix},
\end{align}
where
\begin{align}
    \mathcal{I} &\equiv \begin{pmatrix}
        \frac{\partial^2}{\partial \dot{u}_0^2} & \frac{\partial^2}{\partial \dot{u}_0 \partial \dot{u}_1} \\
        \frac{\partial^2}{\partial \dot{u}_1 \partial \dot{u}_0} & \frac{\partial^2}{\partial \dot{u}_1^2}
    \end{pmatrix} I, \nonumber \\
    \mathcal{F}_i &\equiv \frac{1}{L} \frac{\partial L}{\partial u_i} I + \frac{\partial I}{\partial u_i} - \frac{\partial^2 I}{\partial u_j \partial \dot{u}_i} \dot{u}_j - \frac{1}{\rho} \frac{\partial I}{\partial \dot{u}_i} - \frac{\partial L}{\partial u_j} \frac{\partial I}{\partial \dot{u}_i} \frac{\dot{u}_j}{L},
\end{align}
with $i, j = 0, 1$ and the summation over $j$ is implied. 
From this equation of motion, we can explicitly check the observation in the previous section that the DBI action fails to describe the motion of the D2-brane near the true vacuum, as $\mathcal{I}$ vanishes when both endpoints of the D2-brane are on the same brane, e.g., $u_0 < v'$ and $u_1 = 0$. In the actual numerical calculation, we may avoid this singularity due to the smoothing of the edge.

\begin{figure}[t]
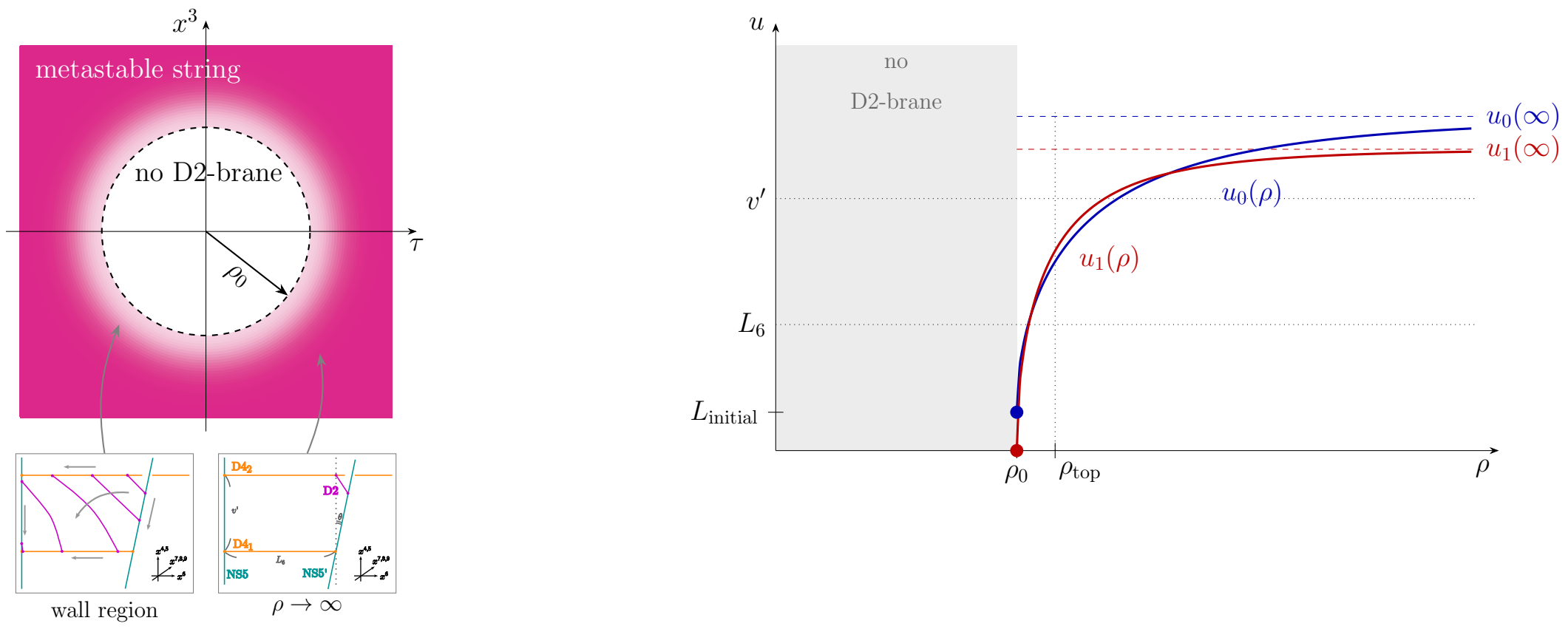

\centering
\includestandalone[width=0.85\textwidth]{tikz/fig_d2_bounce}
\caption{%
    A schematic picture of the conjectured bounce solution for the decay of
    the metastable vortex string, to be compared with
    Fig.~\ref{fig:bounce_scalar}. Left: the $(\tau, x^3)$-plane. In contrast
    to the scalar field case, no D2-brane exists for $\rho < \rho_0$; the
    D2-brane suddenly starts to exist at $\rho = \rho_0$ (dashed circle),
    sweeps the configurations in the lower left inset across the wall region,
    and approaches the metastable vortex string configuration (lower right
    inset) as $\rho \to \infty$. Right: a schematic profile of the endpoint
    parameters $u_0(\rho)$ (blue) and $u_1(\rho)$ (red), which asymptote to
    the false vacuum values (dashed lines); see Fig.~\ref{fig:bounces} for
    the numerical solutions.%
}
\label{fig:bounce_brane}
\end{figure}

Now, we would like to find the bounce solution of the above equation of motion with the overshoot/undershoot method. We need to tune the initial configuration of the D2-brane and $\rho_0$, the time when the D2-brane suddenly appears, to find the solution which approaches the configuration of the D2-brane corresponding to the straight vortex string at $\rho \to \infty$. Then, what is the initial configuration of the D2-brane and which parameters describe the initial configuration? As we mentioned above, the two endpoints of the D2-brane cannot be on the same brane at the initial configuration in our approximation, and we assume that this reflects the fact that the state there is so close to the true vacuum that the DBI action cannot describe it. Therefore, we assume that the initial configuration of the D2-brane is a straight line with length $L_\text{initial}$ whose endpoints are both on the NS5-brane or both on the D4${}_1$-brane; i.e., $u_0 = L_\text{initial}$ and $u_1 = 0$ or $u_0 = 0$ and $u_1 = L_\text{initial}$, as schematically shown in Fig. \ref{fig:bounce_brane}.
The initial configuration of the D2-brane is then described by four parameters, $L_\text{initial}$, $\dot{u}_0$ and $\dot{u}_1$ at $\rho = \rho_0$.

To further simplify the calculation, we assume that the motion of the D2-brane experiences a point where $\mathcal{F}_0 = \mathcal{F}_1 = 0$ at some $\rho = \rho_\text{top}$. We also assume that at $\rho = \rho_\text{top}$, the endpoints of the D2-brane, $X_E(u_0)$ and $X_E(-u_1)$, are on the NS5-brane and the D4${}_1$-brane , respectively. This is motivated by the observation that the bounce solution of the vacuum decay in quantum field theory experiences a point where the potential gradient vanishes, $\partial V / \partial \phi = 0$, at some $r$ when $\phi_\text{bounce}$ is between $\phi_\text{true}$ and $\phi_\text{false}$. We, however, admit that this is just an assumption and we do not have a strong reason to believe that this is the case; for example, we ignore the friction term that may well exist in the equation of motion of the D2-brane just like the friction term in the field theory case, although we expect that the friction term is not important for large $\rho_\text{top}$.
Also, even if our first assumption is the case, the endpoints of the D2-brane may be on the different branes at $\rho = \rho_\text{top}$, e.g., on NS5-brane and NS5$'$-brane, instead of on the NS5-brane and the D4${}_1$-brane. This second assumption is just for numerical convenience as we discuss later, and indeed, we observe that $u_0(\rho_\text{top}) \to v'$ or $u_1(\rho_\text{top}) \to L_6$ as we take smaller $\theta$, and thus the second assumption might be violated in the small $\theta$ limit.
Having said that, as we will see later, we can find the bounce solution with the above assumptions, and we also find that without the above assumptions, we cannot find the bounce solution, which supports the validity of the first assumption to some extent.

With the above assumptions, the initial condition is reduced to three parameters: $u_0(\rho_\text{top})$, $u_1(\rho_\text{top})$, and $\rho_\text{top}$. The velocities $\dot{u}_0(\rho_\text{top})$ and $\dot{u}_1(\rho_\text{top})$ are determined by the condition $\mathcal{F}_0 = \mathcal{F}_1 = 0$. We compute them numerically using the following procedure.

\begin{enumerate}
\item \textbf{Symmetric case:}
For simplicity, we first consider the symmetric case, where $u_0(\rho_\text{top}) = u_1(\rho_\text{top}) \equiv u_\text{sym}$ and we fix $u_\text{sym}$. By symmetry, we have $\dot{u}_0 = \dot{u}_1 \equiv \dot{u}_\text{sym}$ and $\mathcal{F}_0 = \mathcal{F}_1 \equiv \mathcal{F}_\text{sym}$. Since $\mathcal{F}_\text{sym}$ depends only on $\dot{u}_\text{sym}$, we find the value satisfying $\mathcal{F}_\text{sym} = 0$ using the bisection method.

\item \textbf{General case:}
Next, we extend to the general case $u_0(\rho_\text{top}) \neq u_1(\rho_\text{top})$. In this case, we parametrize the configuration by $u_\text{offset} = \min(u_0, u_1)$ and $u_\text{diff} = u_0 - u_1$. For fixed $u_\text{offset}$, we discretize $u_\text{diff}$ into $N_\text{diff}$ steps and compute $\dot{u}_0(\rho_\text{top})$ and $\dot{u}_1(\rho_\text{top})$ for each step. At each step $n$, we minimize $|\mathcal{I}^{-1} \mathcal{F}|^2$ using the Nelder-Mead method.

For the initial simplex in the Nelder-Mead algorithm, we use:
\begin{align}
    &(\dot{u}_0|_{n=k-1} + \delta_0, \dot{u}_1|_{n=k-1}), \nonumber \\
    &(\dot{u}_0|_{n=k-1}, \dot{u}_1|_{n=k-1} + \delta_1), \nonumber \\
    &(\dot{u}_0|_{n=k-1} + \delta_0, \dot{u}_1|_{n=k-1} + \delta_1),
\end{align}
where $\delta_i = \dot{u}_i|_{n=k-1} - \dot{u}_i|_{n=k-2}$ for $k > 2$. For the first step ($k=1$), we use:
\begin{align}
    &(\dot{u}_\text{sym}, \dot{u}_\text{sym}), \nonumber \\
    &(1.00025 \dot{u}_\text{sym}, \dot{u}_\text{sym}), \nonumber \\
    &(\dot{u}_\text{sym}, 1.00025 \dot{u}_\text{sym}),
\end{align}
where $\dot{u}_\text{sym}$ is obtained from the symmetric case\,\footnote{This follows the implementation in \texttt{SciPy}\,\cite{2020SciPy-NMeth}.}.
We verify that $|\mathcal{I}^{-1} \mathcal{F}|^2$ is sufficiently small for each step. Finally, we interpolate the computed values to obtain $\dot{u}_0$ and $\dot{u}_1$ for arbitrary $u_0(\rho_\text{top})$ and $u_1(\rho_\text{top})$.
\end{enumerate}

Our task is then to tune $u_\text{offset}$, $u_\text{diff}$, and $\rho_\text{top}$ to find the bounce solution using a nested bisection approach. The algorithm proceeds as follows:

\begin{enumerate}
\item \textbf{Bisection over $u_\text{diff}$ for fixed $\rho_\text{top}$ and $u_\text{offset}$:}
\label{item:step1}
We first fix $u_\text{offset}$ and $\rho_\text{top} \equiv \rho_\text{trial}$. The fate of the D2-brane depends on $u_\text{diff} = u_0 - u_1$:
\begin{itemize}
    \item When $u_\text{diff}$ is very large and positive, the D2-brane is nearly parallel to the NS5-brane, and the solution overshoots the false vacuum as $X_E(u_0)$ goes beyond the endpoint of the target false vacuum configuration.
    
    \item When $u_\text{diff}$ is very large and negative, the D2-brane is nearly parallel to the D4${}_1$-brane, and again overshoots as $X_E(-u_1)$ goes beyond the endpoint.
    
    \item For intermediate values of $u_\text{diff}$, the solution may either \emph{simultaneously} overshoot (both endpoints together approach and pass the false vacuum endpoints) or undershoot (one endpoint turns back before reaching the false vacuum endpoint). 
\end{itemize}
By bisecting over $u_\text{diff}$, we find the value where the solution transitions between one-sided overshooting and simultaneous overshooting/undershooting regimes, which corresponds to the most-bounce-like solution for that particular $\rho_\text{trial}$.
Note that the optimal $u_\text{diff}$ for the bounce solution is dependent on $\rho_\text{trial}$ in general, since the initial velocities and the equation of motion depend explicitly on $\rho$.

\item \textbf{Bisection over $\rho_\text{top}$:}
\label{item:step2}
Once we have found the most-bounce-like solution for a given $\rho_\text{trial}$ by tuning $u_\text{diff}$, we determine whether this solution simultaneously overshoots or undershoots the false vacuum asymptotically. We assume that small $\rho_\text{top}$ (weak friction) leads to undershooting, while large $\rho_\text{top}$ (strong friction) leads to overshooting, analogous to the vacuum decay problem in field theory. By bisecting over $\rho_\text{top}$, we find the value where the solution transitions from undershooting to overshooting, which identifies the bounce solution. Since the optimal $u_\text{diff}$ depends on $\rho_\text{trial}$, we repeat the bisection over $u_\text{diff}$ for each $\rho_\text{trial}$ during this step.

\item \textbf{Search over $u_\text{offset}$:}
\label{item:step3}
Finally, we repeat the above nested bisection for different values of $u_\text{offset}$. We observe that solutions exist and approach the false vacuum configuration for a wide range of $u_\text{offset}$ values, but the action varies with $u_\text{offset}$. We interpret this as a numerical artifact due to finite precision: ideally, only one specific value of $u_\text{offset}$ corresponds to the true bounce solution, because if we took the initial configuration of the D2-brane at $\rho \to \infty$, we would find that only one specific value of $u_\text{offset}$ leads to the solution that approaches the false vacuum configuration at $\rho \to \infty$, but simply our numerics cannot resolve this. Therefore, we assume that the bounce solution is characterized by the maximum bounce action value across all $u_\text{offset}$ choices, as this most closely resembles the vacuum decay instanton in field theory. Note that, to calculate the bounce action, we solve the equation of motion backwards from $\rho = \rho_\text{top}$ to $\rho = \rho_0$, where either $u_0$ or $u_1$ reaches zero, and then compute the action by integrating from $\rho = \rho_0$ to such $\rho$ that the solution sufficiently approaches the false vacuum configuration.
\end{enumerate}

To illustrate how the above algorithm works, we perform the numerical calculation for some parameters, $L_6 = 6, v' = 25, \ell_s = 20, \theta = 0.1$. 
This choice of parameters satisfies the conditions, Eq.\,\eqref{eq:low_energy_effective_field_theory}, and Eq.\,\eqref{eq:meta-stable_vortex_condition}. Also, we may take, for example, $g_s = 10^{-3}$ so that $g \sim 0.5$.
According to our algorithm, we find $u_\text{offset} = 1.14$, $u_\text{diff} = -0.626$, and $\rho_\text{top} = 57.0$ gives the bounce solution. The bounce action for this solution is $S_\text{inst} = 90.2$ for $g_s = 10^{-3}$.
We note that this value is close to the thin-wall approximation, which gives $R_\text{inst} \sim 60$ and $S_\text{inst} \sim 89.5$ for the above parameters.

\begin{figure}[t]
\centering
\includegraphics[scale=0.65]{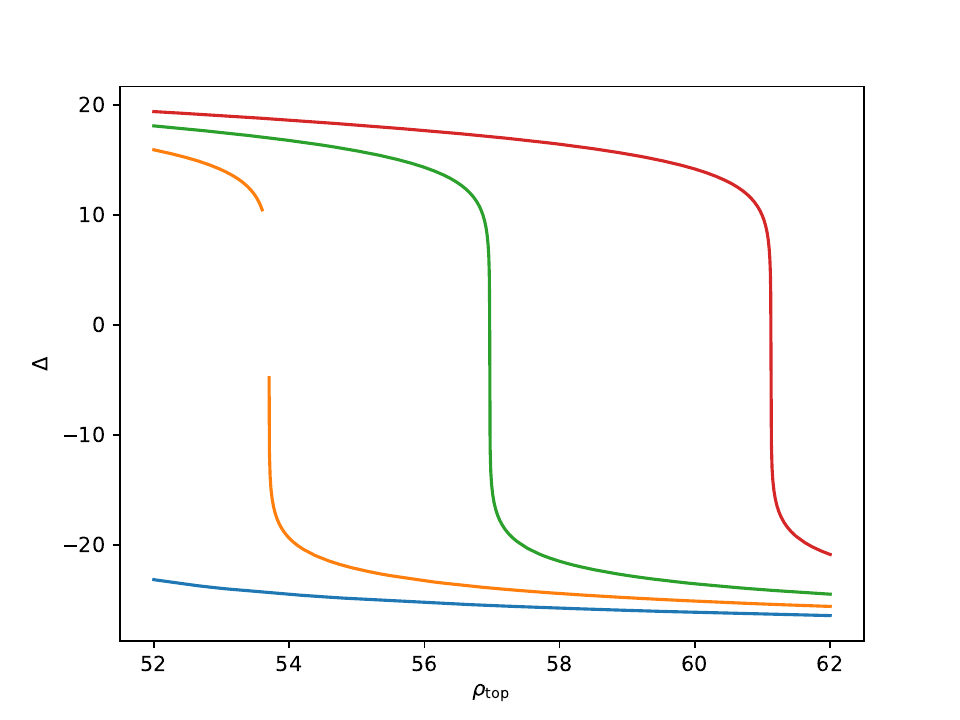}
\caption{
    $\Delta$ as a function of $\rho_\text{top}$ for different values of $u_\text{diff}$ with $u_\text{offset} = 1.14$. The red, green, orange, and blue curves correspond to $u_\text{diff} = -0.55$, $-0.626$, $-0.7$, and $-0.8$, respectively. When the solution is in the undershooting regime, $\Delta$ does not have any value and the curve is therefore interrupted. 
}
\label{fig:rho_udiff}
\end{figure}

In Fig.\,\ref{fig:rho_udiff}, we plot $\Delta$ (to be defined shortly) as a function of $\rho_\text{top}$ for different values of $u_\text{diff}$ with $u_\text{offset} = 1.14$. The red, green, orange, and blue curves correspond to $u_\text{diff} = -0.55$, $-0.626$, $-0.7$, and $-0.8$, respectively. 
The value $\Delta$ is defined as the difference between either $u_0$ or $u_1$ and the corresponding endpoint of the false vacuum configuration when the other endpoint approaches the false vacuum endpoint. The sign of $\Delta$ is defined such that $\Delta$ is positive when $u_0$ is at the false vacuum endpoint and $\Delta$ is negative when $u_1$ is at the false vacuum endpoint.
For example, if $u_0$ is at the intersection of the NS5-brane and the D4${}_2$-brane when $u_1$ is at the false vacuum endpoint, $\Delta$ is $-L_6$. Another example is that when the simultaneous overshooting occurs, $\Delta = 0$. If the solution is in the undershooting regime, $\Delta$ is defined not to have any value.

From Fig.\,\ref{fig:rho_udiff}, we can see how Step\,\ref{item:step1} and Step\,\ref{item:step2} of our algorithm works. First, let us fix $\rho_\text{top}$ to some smaller value, e.g., $\rho_\text{top} = 53.6$. For this $\rho_\text{top}$, we perform Step\,\ref{item:step1}, the bisection over $u_\text{diff}$. For a smaller $u_\text{diff}$, e.g., $u_\text{diff} = -0.8$, $\Delta$ is negative. On the other hand, for a larger $u_\text{diff}$, e.g., $u_\text{diff} = -0.55$, $\Delta$ is positive. We can find the value of $u_\text{diff}$ where $\Delta$ changes its sign by bisection. However, as the orange curve for $u_\text{diff} = -0.7$ shows, in the intermediate value of $u_\text{diff}$, the solution undershoots the false vacuum. We therefore conclude that the solution with $\rho_\text{top} = 53.6$ is in the undershooting regime in Step\,\ref{item:step2}. Next, we are back to Step\,\ref{item:step1} and perform the bisection over $u_\text{diff}$ for a larger $\rho_\text{top}$, e.g., $\rho_\text{top} = 57.0$. For this $\rho_\text{top}$, we similarly find the value of $u_\text{diff}$ where $\Delta$ changes its sign by bisection. Then, as the green curve for $u_\text{diff} = -0.626$ shows, the solution with $\rho_\text{top} = 57.0$ is in the simultaneous overshooting regime in Step\,\ref{item:step2}. We repeat the above procedure to find the value of $\rho_\text{top}$ where the solution transitions from undershooting to overshooting, which gives the bounce solution. With the above parameters, we find $\rho_\text{top} = 57.0$ gives the bounce solution.

\begin{figure}[t]
\centering
\begin{minipage}{0.5\textwidth}
\centering
\includegraphics[scale=0.5]{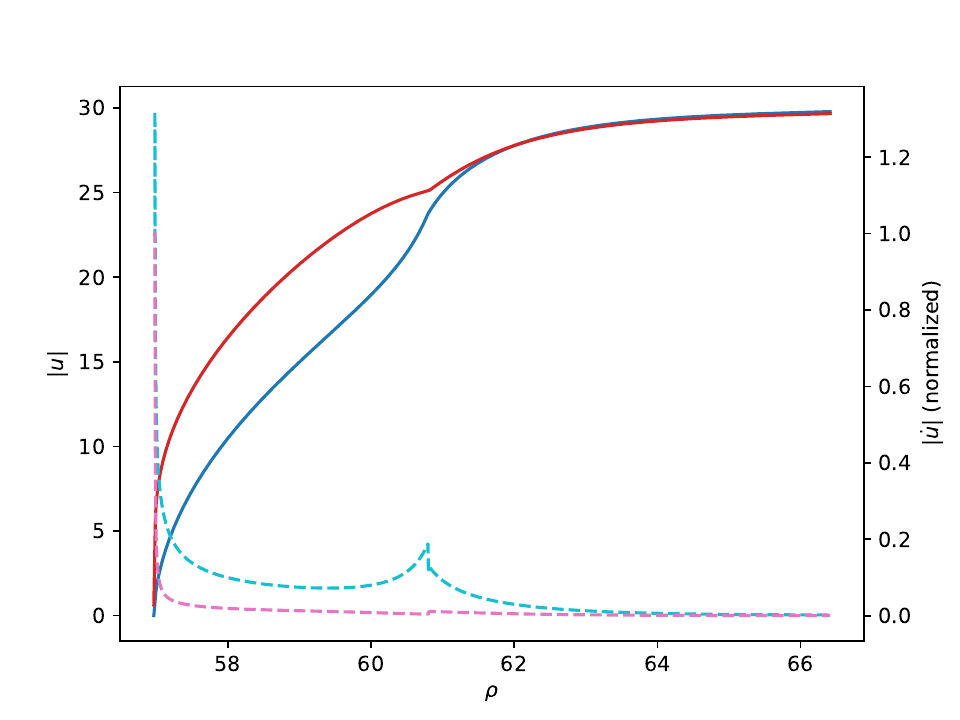}
\end{minipage}\hfill
\begin{minipage}{0.5\textwidth}
\centering
\includegraphics[scale=0.5]{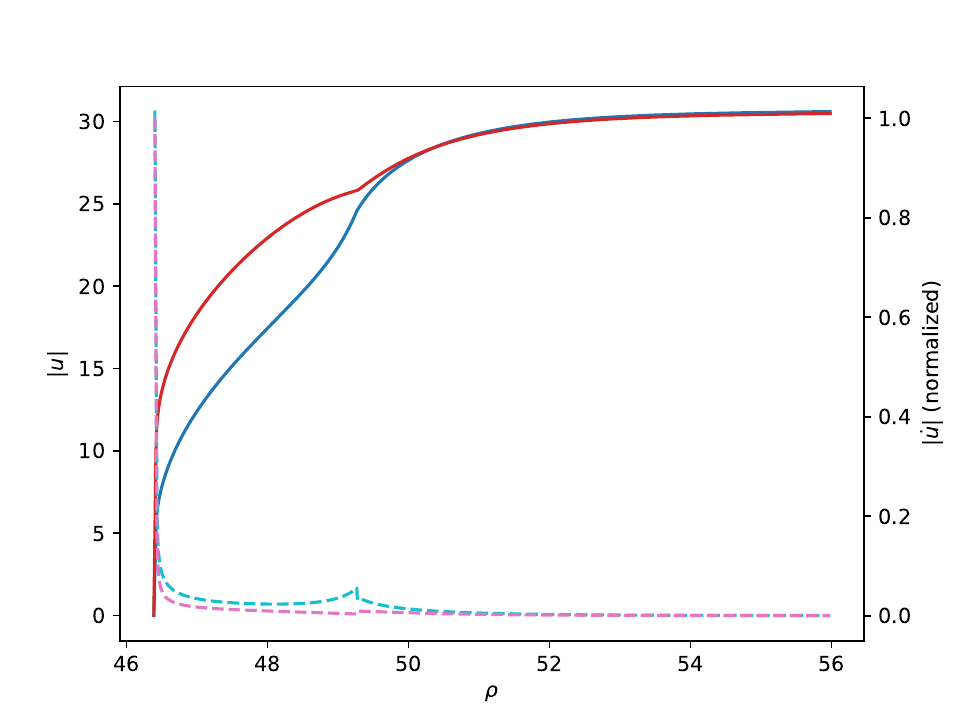}
\end{minipage}
\caption{
    The most-bounce-like solutions for $u_\text{offset}=1.14$ (left) and $u_\text{offset}=0.30$ (right). 
    Solid curves show the endpoint positions: blue for $|u_0(\rho)|$ and red for $|u_1(\rho)|$. 
    Dashed curves show the endpoint velocities: cyan for $|\dot{u}_0(\rho)|/|\dot{u}_0(\rho_\text{top})|$ and pink for $|\dot{u}_1(\rho)|/|\dot{u}_1(\rho_\text{top})|$.
}
\label{fig:bounces}
\end{figure}

To show that Step\,\ref{item:step3} is necessary, we plot the most-bounce-like solutions for $u_\text{offset}=1.14$ and $u_\text{offset}=0.3$ in Fig.\,\ref{fig:bounces}. The left panel shows the solution for $u_\text{offset}=1.14$ and the right panel shows the solution for $u_\text{offset}=0.3$. We identify the former as the bounce solution, as the bounce action for the former, $S_\text{inst} = 90.2$, is larger than that for the latter, $S_\text{inst} = 86.4$. However, as we can see from Fig.\,\ref{fig:bounces}, both solutions approach the false vacuum configuration at large $\rho$, seemingly asymptotically. Within our numerical precision, we cannot determine which solution is the true bounce solution, and thus we assume that the solution with the larger bounce action is the true bounce solution. 

We admit that some numbers in the above calculation are close to or even smaller than the string length $\ell_s$; heavier degrees of freedom such as the stringy excitations may be relevant in this case. Nonetheless, the resulting bounce action is close to the thin-wall result, which suggests that our algorithm captures the decay process at least where the thin-wall approximation is valid. 

\begin{figure}[t]
\centering
\begin{minipage}{0.5\textwidth}
\centering
\includegraphics[scale=0.5]{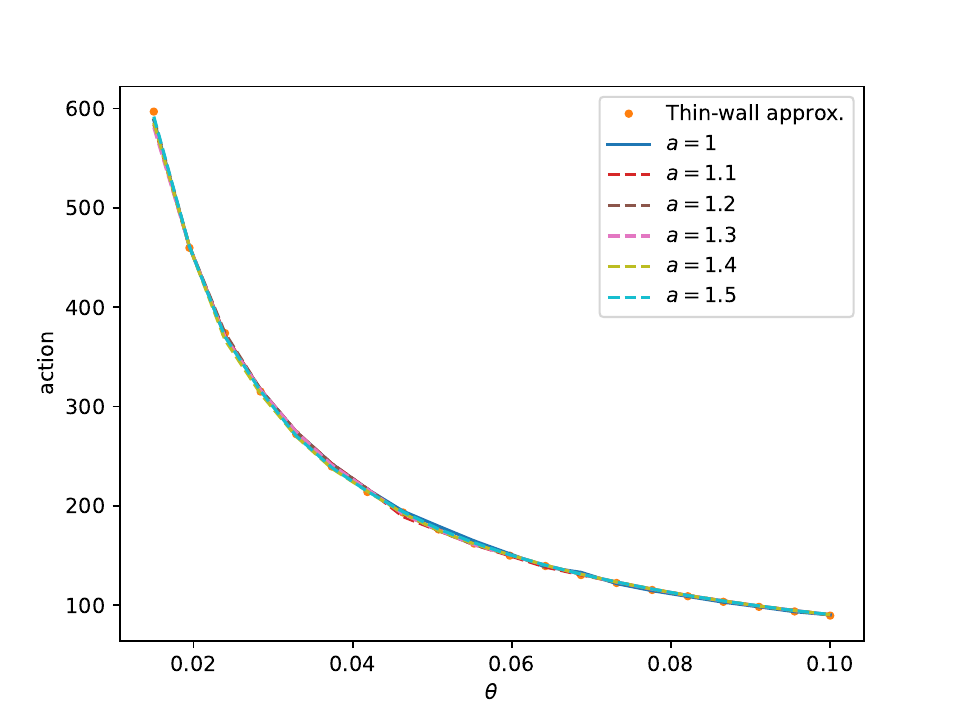}
\end{minipage}\hfill
\begin{minipage}{0.5\textwidth}
\centering
\includegraphics[scale=0.5]{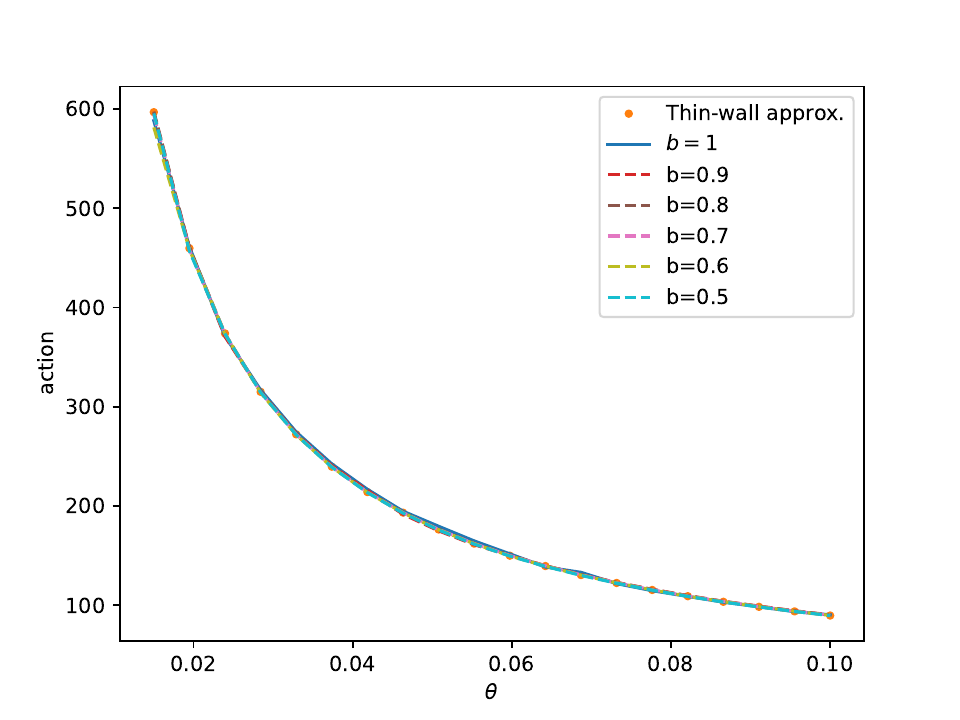}
\end{minipage}
\caption{
    The bounce action for different values of the scaling factors $a$ (left) and $b$ (right). We take $L_6 = 6, v' = 25, \ell_s = 20, g_s = 10^{-3}$ as the reference parameters for $a = b = 1$ and plot the bounce action with varying $\theta$ for different values of $a$ and $b$.
    The correspondence between the scaling factors and the style of the curves are shown in the legends. The bounce action from the thin-wall approximation is also shown as the orange dot for some $\theta$ values for reference. 
    For the left figure, we scale back $\theta$ by $a$ to show the bounce action as a function of the original $\theta$ for different values of $a$.
}
\label{fig:scalings}
\end{figure}

To perform further non-trivial checks of our algorithm, we have also checked the non-trivial scaling of the bounce action with respect to the parameters in the brane picture. That is, different choices of parameters in the brane picture may give the same parameters in the field theory, and thus they should give the same bounce action. 
In addition to the trivial scaling, which scales all the length scales in the brane picture by the same factor,
we have two types of scaling:
\begin{enumerate}
    \item \textbf{Scaling of $g_s$:} $$
    L_6 \to a L_6, \quad g_s \to a g_s, \quad \theta \to a \theta$$
    with $a$ being a scaling factor.
    \item \textbf{Simultaneous scaling of $g_s$ and $\ell_s$:} $$
    \ell_s \to b \ell_s, \quad g_s \to b^{-1} g_s, \quad v' \to b^2 v'$$
    with $b$ being a scaling factor.
\end{enumerate}
For small $\theta$ limit, these scalings does not change the parameters in the field theory, and thus the bounce action should be invariant under these scalings. 
In Fig.\,\ref{fig:scalings}, we plot the bounce action in terms of $\theta$ for different values of the scaling factors $a$ and $b$. Again, we take $L_6 = 6, v' = 25, \ell_s = 20, g_s = 10^{-3}$ as the reference parameters for $a = b = 1$. For the scaling of $a$, we scale back $\theta$ by $a$ to show the bounce action as a function of the original $\theta$ for different values of $a$. As a reference, we also plot the bounce action from the thin-wall approximation; for this calculation, we have used the string tension and monopole mass from the brane picture, $T_\text{vortex} = \tau_2 v' \sin\theta / g_s$ and $m_\text{monopole} = \tau_2 v' L_6 / g_s$, respectively.
We see that all the bounce action curves are sufficiently close to the thin-wall approximation, and the bounce action is indeed invariant under the above scalings.

Before turning to a wider range of $\theta$, let us examine the validity of the thin-wall approximation itself, which the solutions of Fig.\,\ref{fig:bounces} allow us to check directly. In the brane picture, the approximation requires the thickness of the ``wall'', the range $\Delta \rho$ over which the D2-brane configuration deviates appreciably from the metastable one, to be much smaller than the radius of the bubble. This thickness is not fixed by the thin-wall approximation itself; in the wall region the integrand of Eq.\,\eqref{eq:DBI_action_reduced} reduces to $\rho \, |\partial_\rho \vec{X} \wedge \partial_s \vec{X}|$, the density of the area swept by the D2-brane, which is invariant under reparameterizations of $\rho$, so that $\Delta \rho$ is fixed only by the subleading terms. In both panels of Fig.\,\ref{fig:bounces} the D2-brane reaches the metastable configuration within $\Delta \rho \simeq 6$ of its appearance, while the two solutions have different bubble radii, $\rho_\text{top} = 57.0$ and $46.4$. The wall is thus thinner than the bubble by an order of magnitude, and the thin-wall approximation is self-consistent in the brane picture at $\theta = 0.1$.

The same is not true in the field theory picture, where the approximation further requires the bubble to be larger than the core sizes of the monopole and of the vortex string, $1/m_W = 2\pi \ell_s^2 / v' \simeq 100$ and $1/(g q) \simeq 78$, respectively. Both exceed the bubble radius of Eq.\,\eqref{eq:thin_wall_brane_result}, $R_\text{inst} = L_6 / \sin\theta \simeq 60$, for the parameters used here. Equivalently, the monopole core gives the stronger condition, $\sin\theta \ll L_6 v' / (2\pi \ell_s^2) \simeq 0.06$. The thin-wall approximation in the field theory picture is therefore already marginal at $\theta = 0.1$, whereas the one in the brane picture is not.

\begin{figure}[t]
\centering
\includegraphics[scale=0.65]{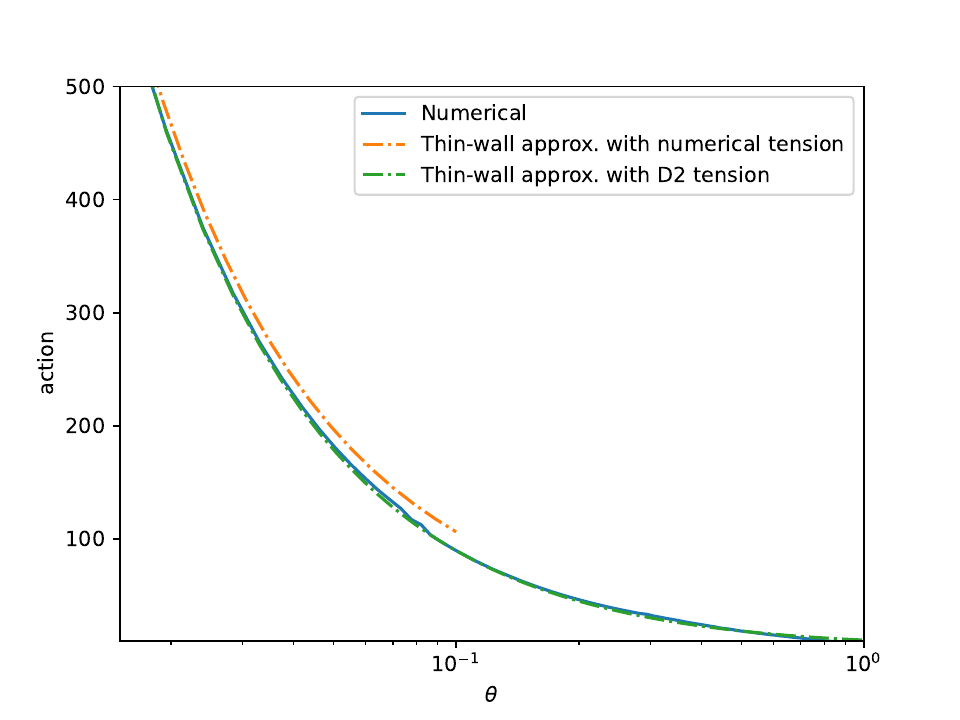}
\caption{
    The bounce action for a wider range of $\theta$. We take $L_6 = 6, v' = 25, \ell_s = 20, g_s = 10^{-3}$ and plot the bounce action by the blue solid line. For the comparison, we also plot the bounce actions from the thin-wall approximation; the orange dash-dot line shows the thin-wall approximation with the string tension from the corresponding field theory and the blue dash-dot line shows the thin-wall approximation with the string tension from the brane picture.
}
\label{fig:brane_comparison}
\end{figure}

Finally, in Fig.\,\ref{fig:brane_comparison}, we plot the bounce action for a wider range of $\theta$ to see if the numerical bounce action deviates from the thin-wall approximation for larger $\theta$. We take $L_6 = 6, v' = 25, \ell_s = 20, g_s = 10^{-3}$ and plot the bounce action by the blue solid line. For the comparison, we plot the bounce actions from the thin-wall approximation with the string tension from the corresponding field theory, which is shown in Fig.\,\ref{fig:tension}, (orange dash-dot line) and with the string tension from the brane picture (blue dash-dot line). For large $\theta$, field theoretic calculation should be unreliable and thus we show the bounce action from the thin-wall approximation with the string tension from the field theory only for $\theta \lesssim 0.1$. 
In the figure, we see that the bounce action from our numerical calculation is close to the thin-wall approximation with the string tension from the brane picture for a wide range of $\theta$.

For $\theta \gtrsim 0.1$, the thin-wall approximation is no longer justified in the field theory picture, as discussed above, while in the brane picture it still is: if $\Delta \rho$ stays of order $L_6$, the ratio of the wall thickness to the bubble radius is $\Delta \rho / \rho_0 \simeq \sin\theta$, and the approximation remains reasonable up to $\theta \sim 0.3$. The agreement with the thin-wall curve in that range is therefore expected from the brane point of view, but it also means that the brane picture does not capture the deviation from the thin-wall result that the field theory bounce should exhibit there. This is the limitation stated in the Introduction: the D2-brane is described as a thin object and does not capture the cores of the monopole and the vortex string, so our description cannot reproduce the deviation of the field theory bounce action from the thin-wall result in this regime.
For $\theta \gtrsim 0.3$, on the other hand, the wall and the bubble become comparable even in the brane picture, and yet
the bounce action from our numerical calculation keeps following the thin-wall curve up to $\theta \sim 1$. We do not have a strong reason for this; 
numerically, our resolution on, in particular, Step\,\ref{item:step3} might not be good enough, where the spacing of $u_\text{offset}$ is taken to be $\sim 0.14$ in our calculation, although we have tried ten times smaller spacing and the bounce action does not change significantly, or it might be because our straight-brane approximation is not good for $\theta \gtrsim 0.3$ and the deviation from the thin-wall approximation is not captured in our numerical calculation.
It is of particular interest to understand the origin of this behavior, which may be a limitation of the straight-brane approximation or of our numerical resolution in $u_\text{offset}$. We leave this as an open question for future work.

\section{Conclusion and Discussion}
\label{sec:conclusion_and_discussion}
In this paper, we have studied the metastable vortex string in the brane picture and its decay process. 
We have calculated and compared the tension of the vortex string in the brane picture and in the field theory, and we have found that they are consistent with each other despite the fact that the D2-brane is a non-BPS object and we have ignored the bending of the branes in our calculation.
We have then derived the bounce action from the DBI action of the D2-brane within the thin-wall approximation, and found that it takes the same form as the field theory one, with the monopole mass given by the area swept by the D2-brane.
As a consistency check, we have developed an algorithm to calculate the bounce action for the decay of the metastable vortex string in the brane picture and performed the numerical calculation based on our algorithm and found that our results successfully reproduce the thin-wall approximation where the thin-wall approximation is expected to be valid.
On the other hand, even in the regime where the thin-wall approximation is not expected to be a good approximation in the field theory, our numerical results are still close to the thin-wall approximation. This is partially expected, as the D2-brane is described as a thin object and does not capture the cores of the monopole and the vortex string, which are responsible for the deviation from the thin-wall result in the field theory.
We have also checked the non-trivial scaling of the bounce action with respect to the parameters in the brane picture, which provides a non-trivial check of our algorithm. In the brane picture, the scaling of $g_s$ and $\ell_s$ with appropriate scaling of other parameters represents non-trivial changes of the parameters, but they do not change the parameters in the field theory, and thus the bounce action should be invariant under these scalings. We have confirmed this invariance of the bounce action under these scalings in our numerical calculation.

The first thing to be discussed is the effect of the non-BPS nature of the D2-brane on the decay process.
Indeed, the non-BPS nature shows up in the tension: Fig.\,\ref{fig:tension} shows that $T_\text{vortex}/T_3$ approaches $1$ as $\theta \to 0$, where the vortex string approaches a BPS one, and deviates from it at finite $\theta$, by about $15\%$ at $\theta = 0.1$. As discussed in Sec.\,\ref{sec:brane_construction}, the deviation comes from the relaxation of the adjoint field in the core of the vortex string, which corresponds to the bending of the D4${}_2$-brane in the $v$ direction. It is interesting to reproduce this deviation in the brane picture by taking into account the bending of the branes, and to see whether higher derivative terms in the effective action of the field theory from our brane configuration, which are not captured in our K\"ahler potential and superpotential, also contribute.

Another way to study the effect of the bending of the branes is to consider a different brane configuration where the BPS limit of the whole brane configuration may be taken; possibly a brane configuration including orientifold planes may be useful for this purpose. One may also consider non-BPS configurations of D4-branes and NS5-branes, such as Ref.\,\cite{Eto:2012kb}, but then it might well be unclear if the correspondence between the brane configuration in Type IIA string theory and the field theory is valid in such a non-BPS configuration.

Once the brane bending effect and/or higher derivative terms in the effective action of field theory from our brane configuration are taken into account, it is interesting to see if the brane picture can reproduce the field theory result for the bounce action even in the regime where the thin-wall approximation is not a good approximation. 
In the field theory picture, several studies have been done to go beyond the thin-wall approximation\,\cite{Shifman:2002yi,Chitose:2023dam,Domcke:2026xza} and it is worth studying if brane picture can reproduce at least some of the results in these studies.

Finally, the description of the bounce by a Euclidean D-brane ending on other branes is not specific to the configuration considered in this paper. It would be interesting to apply it to other brane/field-theory configurations, including those describing the decay of other objects such as domain walls, although it might be more difficult to find the corresponding brane configuration.

\acknowledgments
This work was supported by JSPS KAKENHI Grant Nos.\ 24K17042 [HF], 25H00638 [HF], 26H00403 [HF], 26K17131 [HF] and 24K00522 [KO].
This work was also supported by Grant-in-Aid for JSPS Research Fellow 24KJ0832 [AC]
and by FoPM, WINGS Program, the University of Tokyo~[AC].
This work was also supported by JST PRESTO Grant Nos.\ JPMJPR26AB [HF] and JPMJPR26A3 [KO].
This work was also supported by RIKEN ECL program [KO].
Finally, this work was also supported by the Simons Foundation Grant \#888984 [KO].
In this research work, we used the UTokyo Azure (\url{https://utelecon.adm.u-tokyo.ac.jp/en/research_computing/utokyo_azure/}).

\appendix
\section{String solution in the field theory picture}
\label{sec:vortex_solution}
In this appendix, we introduce the vortex solution in the field theory picture for our superpotential, Eq.\,\eqref{eq:superpotential_full}, and K\"ahler potential, Eq.\,\eqref{eq:kahler_potential_Q}. First, the $F$-term potential is given by
\begin{align}
    V_F &= \frac{1}{2}\text{Tr} \qty(\qty| \frac{\sqrt{2}g}{\cos\theta} Q\tilde{Q} - 2\mu \Phi|^2) + \qty| \tilde{Q} \qty(\frac{\sqrt{2}g}{\cos\theta}\Phi - \frac{m}{\cos\theta} \mathbf{1})|^2 + \qty|\qty(\frac{\sqrt{2}g}{\cos\theta} \Phi - \frac{m}{\cos\theta} \mathbf{1}) Q|^2. 
\end{align}
For the vortex solution, we take the following ansatz:
\begin{align}
    Q = \tilde{Q}^\dagger = \begin{pmatrix}
        q \xi(r) e^{i\varphi} \\ 0
    \end{pmatrix}, \quad \Phi = \frac{v'}{2\sqrt{2}\pi \ell_s^2g}
    \begin{pmatrix}
        \eta(r) & 0 \\
        0 & 0
    \end{pmatrix}, \quad A_\varphi^3 = A_\varphi = \frac{1 - f(r)}{r}
\end{align}
with the following boundary conditions:
\begin{align}
    &\xi(0) = 0, \quad \xi(\infty) = 1, \nonumber \\
    &\eta'(0) = 0, \quad \eta(\infty) = 1, \nonumber \\
    &f(0) = 1, \quad f(\infty) = 0.
\end{align}
With this ansatz, the $D$-term potential is automatically vanishing, and the tension functional is given by
\begin{align}
    T
=
2\pi\int_0^\infty dr\, r & 
\left[
2q^2(\xi')^2
+
2\frac{q^2\xi^2 f^2}{r^2}
+
\frac{1}{g^2}\frac{(f')^2}{r^2}
\right. \nonumber \\ &+ \left.
\qty(\frac{v'}{2\pi\ell_s^2g})^2(\eta')^2
+
\frac{g^2q^4}{\cos^2\theta}(\xi^2-\eta)^2
+
2\frac{m^2q^2}{\cos^2\theta}(\eta-1)^2\xi^2
\right].
\end{align}
Here, $r$ is the radial coordinate in the $x^1$-$x^2$ plane.
The profile functions $\xi(r)$, $\eta(r)$, and $f(r)$ are determined by minimizing the above tension functional. To derive Fig.\,\ref{fig:tension}, we numerically minimize the above tension functional by discretizing the profile functions in $r$ direction.

\bibliography{papers}

\end{document}

%% file: tikz/fig_scalar_bounce.tex
\begin{tikzpicture}[>=Stealth, font=\large]

\begin{scope}[xshift=9.2cm, yshift=-1.55cm, local bounding box=rightpanel]
  \def\phitrue{0.5}
  \def\phizero{0.95}
  \def\phifalse{3.0}
  \def\Rw{3.0}
  \def\ww{0.45}

  \draw[->] (0,0) -- (6.4,0) node[below left=1pt] {$r$};
  \draw[->] (0,0) -- (0,3.9) node[left=2pt] {$\phi$};

  \foreach \yy/\lab in {\phitrue/{\phi_\text{true}}, \phizero/{\phi_0}, \phifalse/{\phi_\text{false}}}{
    \draw (0.07,\yy) -- (-0.07,\yy) node[left=1pt] {$\lab$};
  }

  \draw[dashed] (0,\phifalse) -- (6.2,\phifalse);

  \draw[thin, gray, domain=0:6.1, samples=140, smooth]
    plot (\x, {2.0 - (2.0-\phizero-0.12)*cos(deg(0.95*\x))*exp(-0.13*\x)});

  \draw[thin, gray, domain=0:5.9, samples=140, smooth]
    plot (\x, {\phifalse + (\phizero-0.13-\phifalse)*(1-tanh((\x-\Rw-0.55)/\ww))/2
               + 0.30*pow(max(0,\x-4.55),1.6)});

  \draw[very thick, domain=0:6.1, samples=140, smooth]
    plot (\x, {\phifalse + (\phizero-\phifalse)*(1-tanh((\x-\Rw)/\ww))/2});

\end{scope}

\begin{scope}[local bounding box=leftpanel]
  \def\Rb{2.1}   
  \def\wb{0.32}  
  \def\Lb{3.4}   

  \colorlet{outcol}{blue!55!cyan!80!black}

  \begin{scope}
    \clip (-\Lb,-\Lb) rectangle (\Lb,\Lb);
    \fill[outcol] (-\Lb,-\Lb) rectangle (\Lb,\Lb);
    \foreach \rr in {4.80,4.75,...,0.05}{
      \pgfmathsetmacro{\pc}{max(0,min(100,100*(1+tanh((\rr-\Rb)/\wb))/2))}
      \fill[outcol!\pc!white] (0,0) circle (\rr);
    }
  \end{scope}

  \draw[->] (-\Lb-0.25,0) -- (\Lb+0.45,0) node[below] {$x^{0,1}$};
  \draw[->] (0,-\Lb-0.25) -- (0,\Lb+0.45) node[left] {$x^{2,3}$};

  \fill (0,0) circle (0.075);
  \node[anchor=south west] at (0.12,0.10) {$\phi=\phi_0$};
  \node[white] at (-2.35,2.75) {$\phi\approx\phi_\text{false}$};
\end{scope}

\end{tikzpicture}

%% file: tikz/fig_d2_bounce.tex
\begin{tikzpicture}[>=Stealth, font=\large]

\begin{scope}[local bounding box=diskpanel]
  \def\Rzero{1.9}   
  \def\Rwall{2.35}  
  \def\wb{0.28}     
  \def\Lb{3.4}      

  \colorlet{dtwocol}{magenta!85!violet}

  \begin{scope}
    \clip (-\Lb,-\Lb) rectangle (\Lb,\Lb);
    \fill[dtwocol] (-\Lb,-\Lb) rectangle (\Lb,\Lb);
    \foreach \rr in {4.80,4.75,...,0.05}{
      \pgfmathsetmacro{\pc}{max(0,min(100, 18 + 82*(1+tanh((\rr-\Rwall)/\wb))/2 ))}
      \fill[dtwocol!\pc!white] (0,0) circle (\rr);
    }
    \fill[white] (0,0) circle (\Rzero);
  \end{scope}

  \draw[dashed, thick] (0,0) circle (\Rzero);

  \draw[->] (-\Lb-0.25,0) -- (\Lb+0.45,0) node[below] {$\tau$};
  \draw[->] (0,-\Lb-0.25) -- (0,\Lb+0.45) node[left] {$x^{3}$};

  \draw[->, thick] (0,0) -- (-38:\Rzero) node[midway, below, sloped] {$\rho_0$};

  \node[align=center] at (0.05,1.1) {no D2-brane};
  \node[white, anchor=north west] at (-3.25,3.3) {metastable string};
\end{scope}

\node[draw=gray, thin, inner sep=2pt, fill=white, anchor=north]
  (insetwall) at (-1.85,-4.05)
  {\includegraphics[width=3.1cm]{\figspath/hanany-witten_string_end.pdf}};
\node[anchor=north, font=\normalsize, align=center] at (insetwall.south)
  {wall region};

\node[draw=gray, thin, inner sep=2pt, fill=white, anchor=north]
  (insetfv) at (1.85,-4.05)
  {\includegraphics[width=3.1cm]{\figspath/hanany-witten.pdf}};
\node[anchor=north, font=\normalsize, align=center] at (insetfv.south)
  {$\rho \to \infty$};

\draw[->, gray, thick] (insetwall.north) to[bend left=15] (-133:2.32);
\draw[->, gray, thick] (insetfv.north)  to[bend right=15] (-47:3.05);

\begin{scope}[scale=2, xshift=5.2cm, yshift=-2.0cm, local bounding box=graphpanel]
  \def\rzero{2.2}     
  \def\rtop{2.55}     
  \def\Linit{0.35}    
  \def\uFVa{3.05}     
  \def\uFVb{2.75}     
  \def\vprime{2.3}    
  \def\Lsix{1.15}     

  \fill[gray!15] (0,0) rectangle (\rzero,3.7);
  \node[gray!80!black, align=center, font=\normalsize] at (\rzero/2,3.35)
    {no\\ D2-brane};

  \draw[->] (0,0) -- (6.6,0) node[below left=1pt] {$\rho$};
  \draw[->] (0,0) -- (0,3.9) node[left=2pt] {$u$};

  \draw (\rzero,0.07) -- (\rzero,-0.07) node[below] {$\rho_0$};
  \draw[dotted] (\rtop,0) -- (\rtop,3.1);
  \draw (\rtop,0.07) -- (\rtop,-0.07) node[below right=-2pt] {$\rho_\text{top}$};

  \draw (0.07,\Linit) -- (-0.07,\Linit) node[left=1pt] {$L_\text{initial}$};

  \draw[dotted] (0,\vprime) -- (6.4,\vprime);
  \node[left=1pt] at (0,\vprime) {$v^{\prime}$};
  \draw[dotted] (0,\Lsix) -- (6.4,\Lsix);
  \node[left=1pt] at (0,\Lsix) {$L_6$};

  \draw[dashed, blue!70!black] (\rzero,\uFVa) -- (6.4,\uFVa)
    node[right=1pt] {$u_0(\infty)$};
  \draw[dashed, red!75!black]  (\rzero,\uFVb) -- (6.4,\uFVb)
    node[right=1pt] {$u_1(\infty)$};

  \draw[very thick, blue!70!black, domain=\rzero:6.35, samples=160, smooth]
    plot (\x, {\Linit + (\uFVa-\Linit)*tanh(0.95*sqrt(\x-\rzero))});
  \draw[very thick, red!75!black, domain=\rzero:6.35, samples=160, smooth]
    plot (\x, {\uFVb*tanh(1.35*sqrt(\x-\rzero))});

  \fill[blue!70!black] (\rzero,\Linit) circle (0.06);
  \fill[red!75!black]  (\rzero,0)      circle (0.06);

  \node[blue!70!black] at (4.35,2.35) {$u_0(\rho)$};
  \node[red!75!black]  at (3.05,1.75) {$u_1(\rho)$};
\end{scope}

\end{tikzpicture}